\documentclass[sigconf,nonacm]{acmart}
\usepackage{booktabs} %
\usepackage{multirow}
\usepackage{diagbox}
\usepackage[ruled]{algorithm2e} %
\usepackage{rotating}
\usepackage{gensymb}
\usepackage{amsmath}

\newcommand{\papername}{NeRFFaceEditing}

\SetAlFnt{\small}
\SetAlCapFnt{\small}
\SetAlCapNameFnt{\small}
\SetAlCapHSkip{0pt}

\title{\papername: Disentangled Face Editing in Neural Radiance Fields}

\begin{document}

\author{Kaiwen Jiang} 
\affiliation{%
 \institution{Beijing Key Laboratory of Mobile Computing and Pervasive Device, Institute of Computing Technology, Chinese Academy of Sciences and Beijing Jiaotong University}
\country{China}
}
\email{kevinjiangedu@gmail.com}

\author{Shu-Yu Chen}
\affiliation{%
 \institution{Beijing Key Laboratory of Mobile Computing and Pervasive Device, Institute of Computing Technology, Chinese Academy of Sciences}
 \country{China}
 }
\email{chenshuyu@ict.ac.cn}

\author{Feng-Lin Liu} 
\affiliation{%
 \institution{Beijing Key Laboratory of Mobile Computing and Pervasive Device,Institute of Computing Technology, CAS and University of Chinese Academy of Sciences }
\country{China}
}
\email{liufenglin21s@ict.ac.cn}

\author{Hongbo Fu}
\affiliation{%
\institution{School of Creative Media,\\ City University of Hong Kong}
\country{China}
}
\email{hongbofu@cityu.edu.hk}
 
\author{Lin Gao}
\authornote{Corresponding author is Lin Gao (gaolin@ict.ac.cn).}
\affiliation{%
\institution{Beijing Key Laboratory of Mobile Computing and Pervasive Device, ICT, CAS and University of Chinese Academy of Sciences }
\country{China}
}
\email{gaolin@ict.ac.cn}

\authorsaddresses{Authors' addresses: Kaiwen Jiang, Shu-Yu Chen, Feng-Lin Liu, and Lin Gao are with the Beijing Key Laboratory of Mobile Computing and Pervasive Device, Institute of Computing Technology, Chinese Academy of Sciences. Hongbo Fu is with the School of Creative Media, City University of Hong Kong.  
Authors' e-mails: kevinjiangedu@gmail.com, chenshuyu@ict.ac.cn, liufenglin21s@ict.ac.cn, hongbofu@cityu.edu.hk, gaolin@ict.ac.cn.}

\begin{teaserfigure}
  \centering
  \includegraphics[width=1\linewidth]{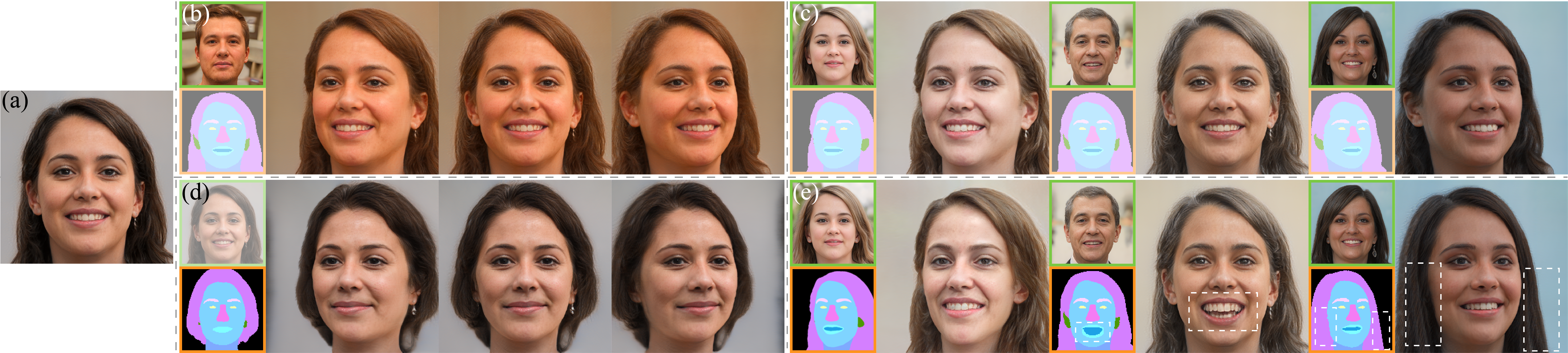}
  \caption{
  Our \papername~method allows users to intuitively edit a facial volume to manipulate its geometry and appearance guided by rendered semantic masks. Given {an input} sample {(a)}, our method disentangles its geometry and appearance, and allows for {one-} or {multi-label} editing. We show a range of flexible face editing tasks that can be achieved with our unified framework: (b) changing the appearance according to a given reference sample while retaining the geometry and 3D consistency; (c) changing the appearance for different views with different reference samples while retaining the geometry; {(d)} editing {multiple labels of} the semantic mask for a certain view while keeping the appearance and 3D consistency; {(e)} editing both the geometry and appearance. The inputs used to control the appearance and geometry are highlighted in green and orange boxes, respectively.}
  \label{fig:teaser}
\end{teaserfigure}

\begin{abstract}
Recent methods for synthesizing 3D-aware face images 
have achieved rapid development thanks to neural radiance fields, allowing for high quality and fast inference speed. 
However, existing solutions for {editing facial geometry} and appearance independently usually require 
retraining and are not optimized for the {recent} work of generation, thus tending to lag behind the generation process.
To address these issues, we introduce \papername, which enables editing and decoupling geometry and appearance in the pretrained {tri-plane-based} neural radiance field while retaining its high quality and fast inference speed. Our key idea for disentanglement is to use the statistics of the tri-plane to represent the {high-level} appearance of its corresponding facial volume. Moreover, we leverage a generated 3D-continuous semantic mask as an intermediary for geometry editing. We devise a geometry decoder (whose output is unchanged when the appearance changes) and an appearance decoder. The geometry decoder aligns the original facial volume with the semantic mask volume. We also enhance the disentanglement by explicitly regularizing rendered images with the same appearance but different geometry to be similar in terms of color distribution for each facial component separately.
Our method allows users to edit via semantic masks with decoupled control of geometry and appearance. Both qualitative and quantitative evaluations show the superior geometry and appearance control abilities of our method compared to existing and alternative solutions.
\end{abstract}



\maketitle

\section{Introduction}
Efficiently generating consistent and high-quality 3D-aware face images is an active research topic.
Many recent techniques (e.g., \cite{gu2021stylenerf, zhou2021cips, or2021stylesdf, deng2022gram, chan2021efficient})
choose to 
build upon Generative Adversarial  \cite{goodfellow2014generative} Neural Radiance Fields (NeRF) \cite{mildenhall2020nerf}{ and form facial volumes to generate 3D-aware high-resolution face images {with 2D convolution.}
}
However, {their methods} lack direct control of facial geometry %
{and some of them {(e.g., \cite{chan2021efficient})}
cannot control the appearance independently of the geometry}, while such control{s} 
{are} important for applications like 3D character design, educational training, {etc.}

{To directly control the geometry,} 
one approach is to introduce an editing intermediary {that is} %
aligned with the facial volume. %
{Semantic masks are suitable for 3D GANs because of their intuitiveness, ease of use, and continuity {while moving the camera}.}
FENeRF \cite{sun2021fenerf} has been proposed 
based on \(\pi\)-GAN 
to generate a volume where facial semantics and texture are spatially aligned. 
However, FENeRF requires time-consuming and resource-hungry retraining {for enabling local editing}. %

On the other hand, 
in NeRF, to control the appearance independently {of the geometry}, %
one 
approach is to directly incorporate the latent code of appearance into a separated branch 
of color 
\cite{schwarz2020graf, niemeyer2021giraffe, liu2021editing, jang2021codenerf, sun2021fenerf}{. This idea} is proven to be effective {and} 
mostly applied in the coordinate-based MLP representation.
Recently, many new representations have been proposed {and particularly}, %
{tri-plane-based radiance field{s}}
proposed in \shortcite{chan2021efficient}
{feature} %
finer details, better quality, and faster inference {than the coordinate-based MLP representation}. %
However, {despite {the advantages of the tri-plane representation}}, %
how to decouple geometry and appearance in {it} 
while preserving {its} 
characteristics remains {unexplored}. 
Trivially extending methods working on the coordinate-based MLP representation to 
{it} 
might abandon its efficient tri-plane representation.
{Actually, there are abundant 2D style control methods (e.g., \cite{huang2017arbitrary, chen2022review}) 
, but their extension {to} 3D generation {tasks} remains an open problem.} %

In this work, we {introduce \papername, which relies on}
{the pretrained tri-plane representation} \cite{chan2021efficient} %
to address {the} above-mentioned limitations {and aims to achieve} 
better {frontal- and side-view editing} and disentanglement of geometry and appearance.
In order to disentangle geometry and appearance {in} 
{the} tri-plane representation {and inspired by adaptive instance normalization (AdaIN) \cite{huang2017arbitrary}}, %
we use {the mean and variance of tri-planes}
to represent the high-level \emph{spatial{ly}-invariant} appearance of its corresponding \emph{spatial{ly}-variant} facial volume. Moreover, we take the merits of the idea that colors are predicted through an additional separated branch and {thus} 
{split {the} original decoder in EG3D} %
into {a} geometry decoder {to handle the geometry} and {an} appearance decoder {to handle the appearance}. {The} geometry decoder takes in features sampled from {the} normalized 
tri-plane{, while the appearance decoder takes in features sampled from the denormalized tri-plane}, so that the geometry of {the} facial volume {will not be} affected {when {the} tri-plane is stylized differently for different appearance}.  %
We choose {a generated} 3D-continuous semantic mask as an intermediary to enable {geometry} editing. 
{T}he key enabler {of effective editing} %
is to make {the} geometry decoder directly predict both {the} semantic label{s} and densities to align the facial volume with the semantic mask volume.

{To enhance} 
disentanglement, {we design a training strategy.}
We utilize a histogram color loss \cite{afifi2021histogan}
to constrain rendered images{ with {the} same appearance but different geometry} to be similar in terms of color distribution for each facial component.

{As shown in Fig. \ref{fig:teaser}, 
\papername~allows} disentangled control of geometry and appearance for better {frontal- and side-view} editing, enabled by the shared geometry space between {the facial volume and the semantic mask volumes}.
{Qualitative and quantitative experiments show that our approach} outperforms state-of-the-art methods for various applications. To facilitate further {research} studies, we will release our code.

The main contributions are summarized as follows:
\begin{itemize}
    \item We propose an AdaIN-based method{ and a design of decoders} to decouple geometry and appearance embedded in {the} {tri-plane} {and enable intuitive geometry editing by semantic masks}.
    \item We propose a fine{-}tuning method to facilitate disentanglement by promoting similarities in terms of color distribution for each facial component separately in rendered images with {the} same appearance but different geometry.
    \item Our method achieves state-of-the-art 3D-aware {frontal- and side-view} editing based on semantic masks as well as {the} disentanglement of geometry and appearance proven by extensive experiments and comparisons.
\end{itemize}

\begin{figure*}[tbh]
    \renewcommand\tabcolsep{0.0pt}
    \renewcommand{\arraystretch}{0}
    \centering \small
    
    \begin{tabular}{c}
        \includegraphics[width=0.9\linewidth]{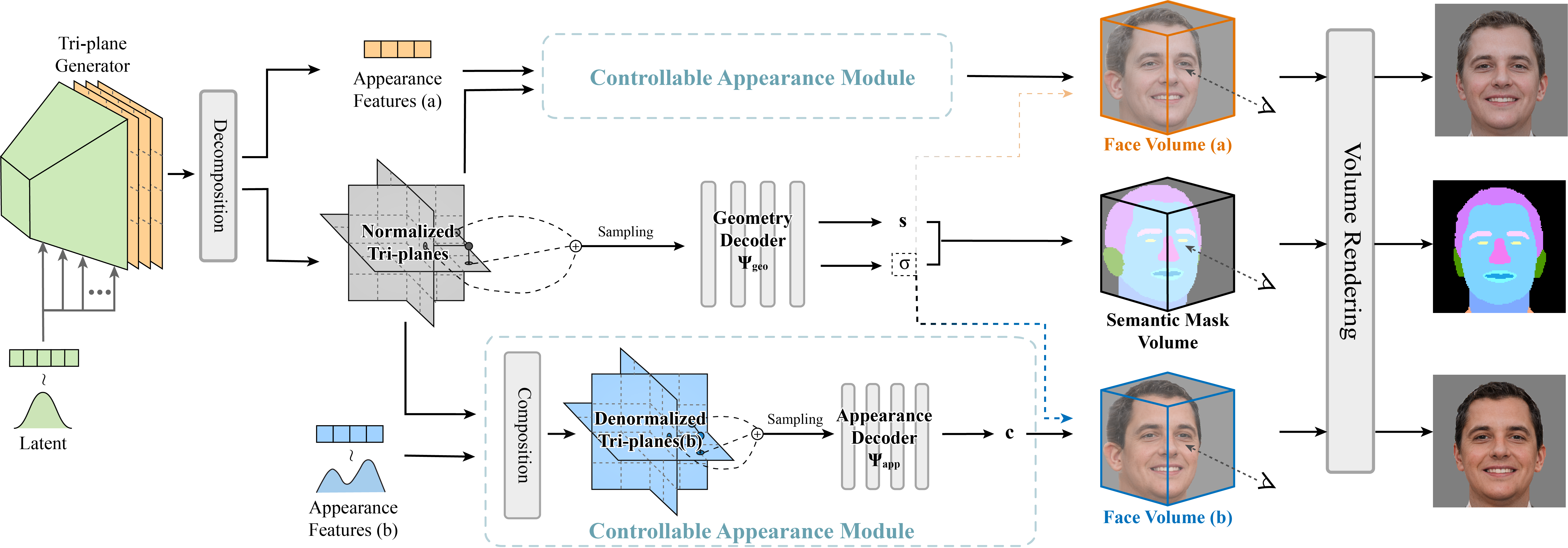}
    \end{tabular}
    \caption{
    An overview of our framework. Our pipeline leverages {the pretrained tri-plane generator to synthesize feature images.}
    {The generated {feature images}
    are then {decomposed into normalized tri-planes and %
    appearance features (a) $F_{\text{app}}^{(a)}$ through reshaping and normalization.}
    } 
    The {geometry} features sampled from {the} normalized tri-plane{s} are passed into {a} geometry decoder to output {densities} \(\sigma\) and semantic label{s} \(\mathbf{s}\), which {together} %
    generate {a} semantic mask volume{ that} {2D semantic masks are projected from. 
    }
    {The normalized tri-planes{, together with appearance features (a) $F_{\text{app}}^{(a)}$ and appearance features (b) $F_{\text{app}}^{(b)}$}, are passed into the Controllable Appearance Module (CAM) to composite the denormalized tri-planes (a) {(not shown in the figure for brevity)} and (b), 
    respectively.}
    Features sampled from {the denormalized} tri-plane{s} (a) and (b) are then passed into {the same}
    appearance decoder to output {color features}
    \(\mathbf{c}\) 
    {in the CAM}{. These color features together with the same densities $\sigma$ are then processed independently by a neural volume renderer to project their corresponding face volumes (a) and (b) into 2D feature images.}}
    \vspace{-3mm}
    \label{fig:overview-of-framework}
\end{figure*}

\section{Related Work}
Our work is closely related to several topics, including disentangled neural implicit representations, 3D-aware neural face image synthesis, {and} neural face image editing.

\subsection{Disentangled Neural Implicit Representations}
Neural implicit scene representation is an emerging area of study, {and}
is originally modelled as an MLP mapping from positional-encoded coordinates to densities and colors for volume rendering.
Recent developments propose to use alternative representations, including hashing table{s} \cite{mueller2022instant}, octrees \cite{yu2021plenoctrees}, {voxels \cite{yu_and_fridovichkeil2021plenoxels},} tri-plane{s} \cite{chan2021efficient}, etc. These methods feature faster inference and improved expressive power.

Given a dataset of single-view 2D {facial} images, %
to have a {disentangled}
control of geometry and appearance, one option is to embed separated latent codes for geometry and appearance into generation {(e.g., \cite{schwarz2020graf, niemeyer2021giraffe, chan2021pi, xu2021volumegan}), }which{, however, does not {directly} provide intuitive {geometry} manipulation}.
{Our solution is to {utilize the tri-plane representation, whose geometry and appearance are decoupled through an AdaIN-based method} {and} {the} original decoder in EG3D is split into {an appearance decoder and} a geometry decoder. The {geometry decoder} 
models both {the} facial volume and the semantic mask volume{, with the latter providing} an editing interface. 
Our solution further enhance{s} the disentanglement through a training strategy.}

\subsection{3D-aware Neural Face Image Synthesis}
In recent years, generative models like Generative Adversarial Networks (GANs) \cite{goodfellow2014generative} combined with implicit radiance fields \cite{mildenhall2020nerf} have been explored to generate 3D-consistent faces {from 2D images only}. 
{To generate high-resolution images, a group of methods {\cite{niemeyer2021giraffe, gu2021stylenerf, zhou2021cips} first} output low-resolution features {and then pass them} %
into 2D convolution}. 
However, 
{they }suffer from the %
low-quality geometry representations.
Thus, alternative models {and representations (e.g., \cite{or2021stylesdf, deng2022gram, chan2021efficient})} {have also been} explored. 
{These models} can synthesize highly realistic images with geometrically-consistent fine details.

However, %
all {the} above methods cannot control geometry in {an} intuitive manner. In order to address this issue, many works have employed the method of embedding explicit control into the generation {process}. {For example}, 
CG-NeRF \cite{jo2021cg} introduce{s} various soft conditions{, including sketches} as input{.}
FENeRF \cite{sun2021fenerf} involve{s} semantic masks in the generation process as output. However, {the quality of their editing results still {has} %
room for improvement}
{and a retraining is unavoidable. By comparison, }
{Sem2NeRF \cite{chen2022sem2nerf} encodes single-view semantic masks into the latent space of pretrained 3D GANs.}
{Lin et al. \shortcite{lin20223d} introduce the work \cite{abdal2021styleflow} on pretrained 2D GANs into 3D GANs to edit the attributes of generated results semantically.}
IDE-3D \cite{sun2022ide}, which is concurrent {with our} work, achieves {interactive high-quality geometry} %
editing and disentangled appearance control. It designs an encoder to facilitate the editing and splits the original tri-planes into semantic tri-planes and texture tri-planes for disentanglement.
{In contrast, our method extend{s} the pretrained tri-plane-based generative model with our unique design of decoders and operations on tri-planes for enabling intuitive editing and decoupling {of} geometry and appearance. {Our method}
achieve{s} 
better results with another training strategy.}

\subsection{Neural Face Image Editing}
With the introduction of {GANs}, various 2D-based methods and 3D-based methods have been proposed to realize {face image editing} %
under different conditions. 
Here, we 
mainly review the approaches based on 3D generative adversarial neural radiance fields. %

In 3D GANs, 
some methods (e.g., \cite{Gafni20arxiv_NerFace, athar2021flameinnerf, zhuang2021mofanerf, hong2021headnerf, zheng2022imavatar}) borrow latent codes for identity, expression, etc. from 3DMM (e.g. \cite{tran20183dmm, FLAME:SiggraphAsia2017}){.}
Some other methods (e.g., \cite{guo2021adnerf, park2021hypernerf, Park20arxiv_nerfies, kania2022conerf, Wang2021DynamicHumanHeads}) use learned embeddings to capture dynamic facial actions, which{, however, are} hard to interpret except for the correspondence with extra information, such as audio or using an attribute regressor.
{FENeRF \cite{sun2021fenerf} incorporates semantic masks into conditional NeRF \cite{schwarz2020graf, chan2021pi} to generate editing interfaces.}
{Similarly, our work also utilizes semantic mask for %
intuitive editing %
with a 3D GAN as {the}
backbone.  %
However, our method %
carefully preserve{s} its original generation power and speed {without requiring retraining}}, {and enhances the disentanglement of geometry and appearance through an additional training strategy.}

\section{Methodology}
In this section, we formalize the structure of our proposed disentanglement framework in detail, which operates on the tri-plane representation \cite{chan2021efficient}.
To decouple geometry and appearance, inspired by AdaIN \cite{huang2017arbitrary}, we decompose the tri-plane {of each sample} {into the \emph{spatially-invariant} abstract appearance features, namely the mean and variance of the tri-plane, and the normalized tri-plane where \emph{spatially-variant} specific geometry features are sampled from (as shown in Fig. \ref{fig:overview-of-framework}). }
Furthermore, we split the original decoder in EG3D into {an appearance decoder for predicting color features and} a geometry decoder {for} predicting densities and semantic labels to enable semantic-mask guided editing
(Sec. \ref{disentanglement-and-mask-guided-editing}).
To further ensure disentanglement, 
we explicitly regularize rendered images with the same appearance but different geometry to be similar in terms of color distribution for each facial component guided by {the} generated semantic masks (Sec. \ref{training-process}). 
The editing during inference
{{will be} %
explained in Sec. \ref{editing-during-inference}}.

\subsection{Preliminaries}
Since %
{our approach is built on the tri-plane representation
proposed in}
EG3D, it is necessary to briefly summarize {EG3D's} pipeline of generation here.
Three planes (tri-plane {for short}) ($p_{xy}$, $p_{xz}$, $p_{yz}$) are generated by StyleGAN2 \cite{karras2020analyzing} \(f\) from {an} intermediate latent code \(w\in W\){.}
For each queried 3D position \(\mathbf{x} = (x, y, z)\), its corresponding feature vector \((\mathbf{F}_{xy}, \mathbf{F}_{xz}, \mathbf{F}_{yz})\) is retrieved by projecting {$\mathbf{x}$} %
onto each of the three planes via bilinear interpolation, {and} is further aggregated by summation to form final feature{s}. %
An additional light-weight decoder network, implemented as a small MLP \(\Phi\), interprets the aggregated 3D features \(\mathbf{F}({\mathbf{x}})\) as color features \(\mathbf{c}({\mathbf{x}})\) and densities \(\sigma({\mathbf{x}})\).
These quantities are rendered into feature images \(I_F\){,} whose first three channels are extracted as rendered images \(I_{RGB}\) in {a} low resolution using
volume rendering \cite{nmax1995optical, mildenhall2020nerf}. The {feature images} \(I_F\) are %
later passed into {a} super-resolution module{,} %
{which generates
high{-}resolution images \(I^{+}_{RGB}\)}. The details of the super-resolution module and the discriminator are omitted for simplicity since they are not our focus.

\begin{figure}
    \renewcommand\tabcolsep{0.0pt}
    \renewcommand{\arraystretch}{0}
    \centering \small
    
    \begin{tabular}{c}
        \includegraphics[width=\linewidth]{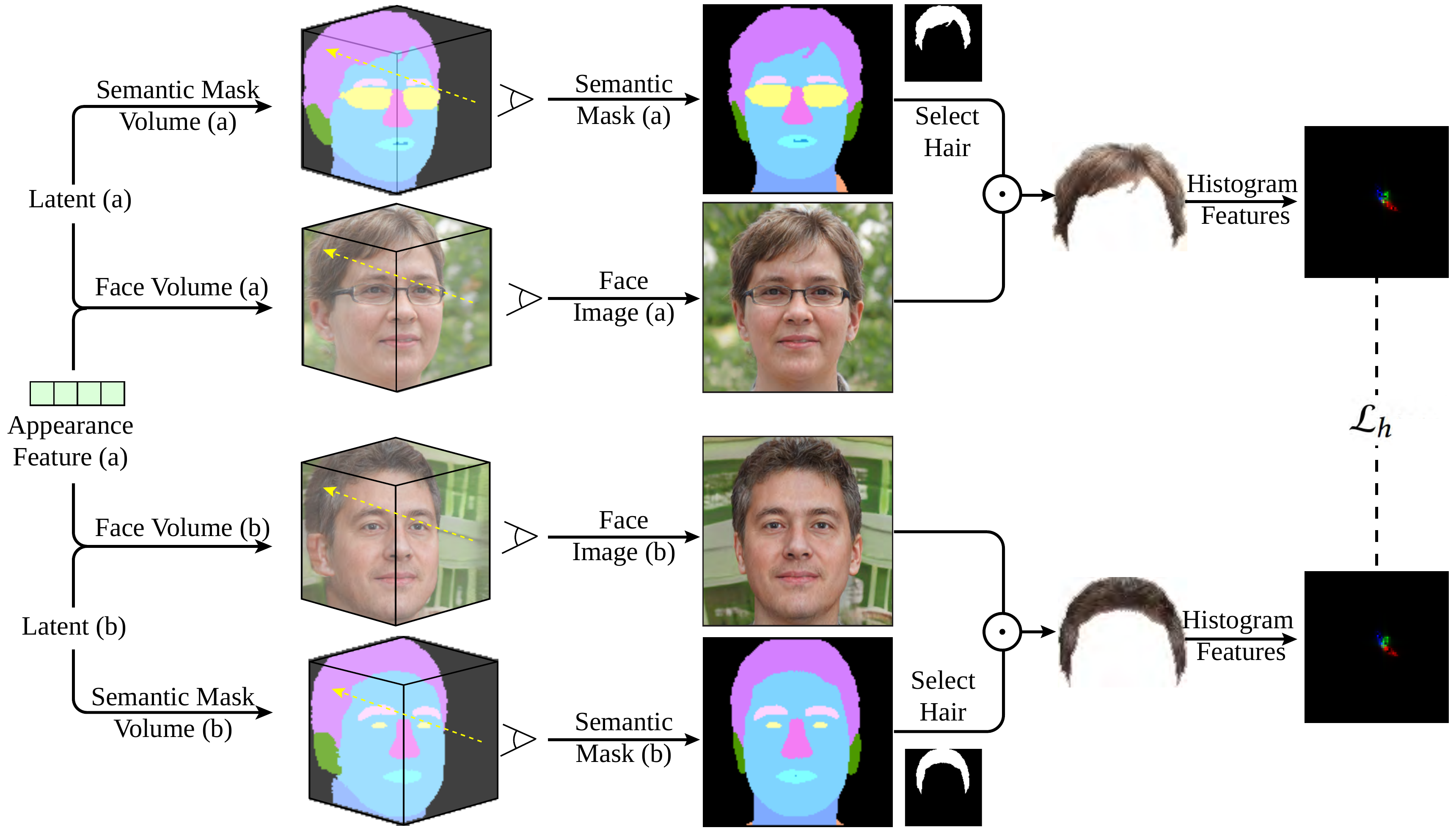}
    \end{tabular}
    
    \caption{Illustration of promoting similarities in terms of color distribution for face images with the {same appearance but different geometry}. For generated face images (a) and (b) at {a certain} pose, {their} similarity is measured by {the} distance between histogram features for each facial component, including hair, eye, lip, skin, etc., separately, {each of} which is selected through {the corresponding generated} semantic mask{s}. In this diagram, we take the hair as an example.}
    \label{fig:promoting-similarity}
\end{figure}
\subsection{Disentanglement and Mask-guided Editing}
\label{disentanglement-and-mask-guided-editing}
Aiming at decoupling the geometry and appearance, 
one possible way is to use {the} cyclic swapping loss in \cite{chen2021deepfaceediting}. Even though it works well in 2D frontal images, it cannot be trivially utilized in the 3D-aware facial {image} generation since there {does not} exist a {consistent} mapping from rendered images at any pose back to {their} unique appearance features. %

{I}t has been known that convolutional feature statistics can capture the style of 2D image{s} \cite{li2016combiningforimagesynthesis, li2017demystifying, leon2016image_style_transfer, huang2017arbitrary}. However, 
{how to apply this idea to 3D-aware {image} generation is challenging,} especially in the context of neural radiance fields where convolutional features usually do not exist.
But in the case of the tri-plane representation, we are able to extend {the conclusion of} AdaIN \cite{huang2017arbitrary} {(i.e., the mean and variance can reflect the style of {a} 2D feature map)} to tri-plane{s}{,} which %
{are} 
practically multi-channel convolutional feature{s}. %
Thus, we assume {that} the mean and variance of the tri-plane \(\mathbf{F}_{app}\) reflect its style{,} which %
represents the high-level appearance of the corresponding facial volume. The obvious benefit is that for a specific latent code \(w\), its representation of appearance is the same for any pose \(P\).

{As in AdaIN, the style of the tri-plane is specifically controlled by the normalization and denormalization operations.}
{We define} the normalization process of {the} %
tri-plane and the denormalization process with any \(\widehat{\mathbf{F}}_{app}\) respectively as:
\begin{equation}
    \overline{p}_{i}=\frac{p_{i}-\mu_{i}}{\sigma_{i}}, \quad  
    p'_{i}=\widehat{\sigma}_{i}\overline{p}_{i}+\widehat{\mu}_{i},
\end{equation}
where \(i\in \{xy, xz, yz\}\), $\mu$, $\sigma$ denote the mean and the variance.

{Based on the assumption that the same geometry can have various different appearances {and taking the merits of the idea that colors are predicted through a separated
branch}}, we split the original decoder \(\Phi\) into a geometry decoder \(\Psi_{\text{geo}}\) {and an appearance decoder {$\Psi_{\text{app}}$}. The former} takes in {the geometry} feature{s} sampled from {the} normalized tri-plane as \(\overline{\mathbf{F}}(\mathbf{x})\) and {the latter} takes in feature{s} sampled from {the} denormalized tri-plane as \({\mathbf{F}}'(\mathbf{x})\).
{We name the process of denormalizing normalized tri-planes with appearance features and decoding features sampled from denormalized tri-planes into color features as Controllable Appearance Module (CAM), as shown in Fig. \ref{fig:overview-of-framework}. Thus, when the CAM is fed with the same normalized tri-planes but different appearance features, the appearance of facial volumes is changed while the geometry is unaffected.}

\begin{figure}[t]
    \renewcommand\tabcolsep{0.0pt}
    \renewcommand{\arraystretch}{0}
    \centering \small
    
    \begin{tabular}{c}
        \includegraphics[width=0.9\linewidth]{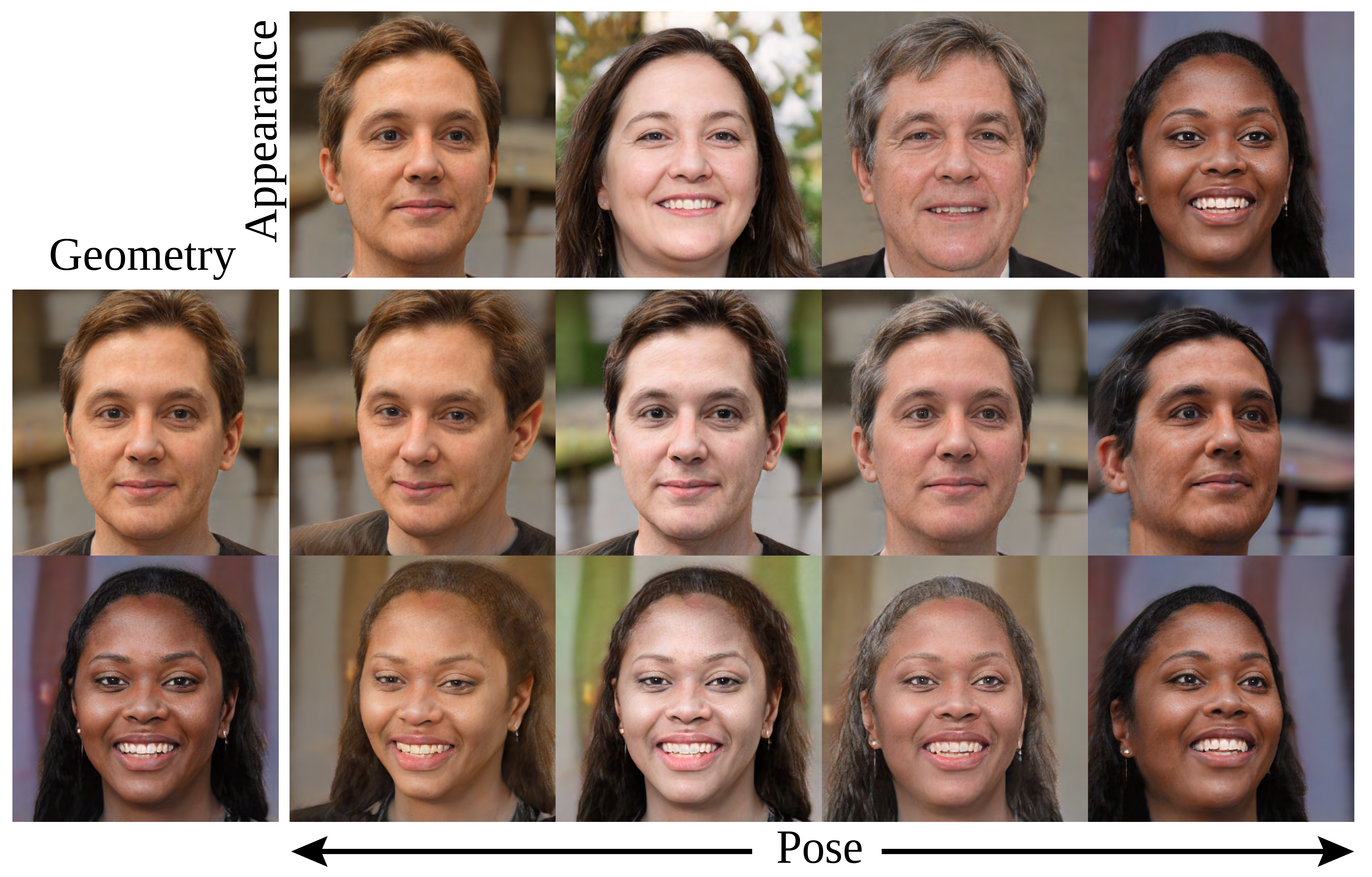}
    \end{tabular}
    
    \caption{Style transfer. The geometry inputs (first column) are the same {as} the appearance inputs (first row). Each face image is generated with the {the geometry reference sample in the same row} and the appearance {reference sample in the same column}. 
    }
    \label{fig:style-transfer-map}
\end{figure}

Moreover, {inspired %
by EditGAN \cite{ling2021editgan}}, {we enable} mask-guided editing in the pretrained uneditable tri-plane representation {based on the following key insight:}
information is concentrated in {the} sampled feature{s} \(\mathbf{F}({\mathbf{x}})\) from {the} tri-plane{,} {and such information concentration} %
is similar to the bottleneck in {an} encoder-decoder architecture. A lightweight decoder is responsible for transforming feature{s} in {an} abstract domain to various specific domains, such as densities and colors, {as well as} other domains {like} semantic labels \(\mathbf{s}\). Thus, we explicitly require the geometry decoder \(\Psi_{\text{geo}}\) to predict both {densities} \(\sigma\) and semantic label{s}. %
{The} above pipeline can be summarized as:
\begin{equation}
    (\sigma(\mathbf{x}), \mathbf{s}(\mathbf{x}))=\Psi_{\text{geo}}(\overline{\mathbf{F}}(\mathbf{x})),  \quad   (\mathbf{c}(\mathbf{x}))=\Psi_{\text{app}}({\mathbf{F}}'(\mathbf{x})),
\end{equation}

However, training another separated decoder which outputs semantic labels \(\mathbf{s}\) only fails to edit precisely
through GAN inversion (see the "Baseline3" of part (a) in Fig. \ref{fig:ablation_study}). We {speculate} that in the case of separated branches for densities \(\sigma\) and semantic labels \(\mathbf{s}\), the spaces between facial volumes and semantic masks are not shared or aligned and the changes in the space of semantic masks fail to propagate to the space of facial volume well. The connection between the space of facial volumes and the space of semantic masks is established through this unified decoder.
Besides, empirically, we find that disentanglement is beneficial for {effective} %
editing %
{of the hair, nose, etc.} (see the "Baseline1" of part (a) of Fig. \ref{fig:ablation_study}).

\subsection{Training Process}
\label{training-process}
{To further improve the disentanglement, we design} a specific training strategy. 
For simplicity, we exemplify the training with the case of low-resolution image {generation}. The case of high-resolution image {generation} only needs {an} additional reconstruction loss (see the supplement materials %
for details).
At {each} step, we first 
{sample }{a} latent code \(w\). {Then, we carefully design the following losses to train our decoder{s} $\Psi_{\text{geo}}$ {and} $\Psi_{\text{app}}$ and fine-tune {the} original tri-plane generator $f$ as $\widetilde{f}$.} We denote {the} original generation process as $G(w)$ and our new 3D-aware and disentangled generation process as $\widetilde{G}(w, \mathbf{F}_{\text{app}})$
. The rendering pose is sampled from the pose distribution of {face images in the} dataset and omitted in all {the} following 
equations {for brevity}.
\subsubsection*{Reconstruction Loss.}
To ensure high quality and diverse generation, we need to match the original generated distribution. Specifically, the training procedure is defined as:
\begin{equation}
\begin{aligned}
    (I_{RGB}, d) &= G(w), \quad  (I'_{RGB}, d', S') = \widetilde{G}(w, \mathbf{F}_{app}(w)), \\
    \mathcal{L}_{Recon} &= \lambda_1 || I_{RGB} - I'_{RGB} || + \lambda_2 \mathcal{L}_{VGG}(I_{RGB}, I'_{RGB}) \\
    &+ \lambda_3 E(\Theta(I_{RGB}), S') + \lambda_4 || d - d' ||,
\end{aligned}
\end{equation}
where \(\mathcal{L}_{VGG}\) is {the} perceptual loss introduced in \cite{zhang2018lpips}, which measures the visual similarity between the generated images and {the} input images by a pretrained VGG-19 model, \(d\) and \(d'\) {respectively represent the} ground-truth depth image and {the} reconstructed depth image extracted when performing the volume rendering following \cite{mildenhall2020nerf}, and \(E\) denotes the pixel-wise cross-entropy loss. \(\Theta(\cdot)\) stands for {an} off-the-shelf facial image segmentation module \cite{yu2018bisenet} and \(S'\) represents predicted semantic masks by the geometry decoder \(\Psi_{\text{geo}}\). In our experiments, we empirically set \(\lambda_1=15, \lambda_2=15, \lambda_3=1, \lambda_4=5\).

\begin{figure}[t]
    \renewcommand\tabcolsep{0.0pt}
    \renewcommand{\arraystretch}{0}
    \centering \small
    
    \begin{tabular}{ccccccc}
        
        & Source
        & Mask
        & Modified
        & \multicolumn{3}{c}{Free-viewed Editing}
        \\
        \rotatebox{90}{\hspace{2mm}+Glass} &
        \includegraphics[width=0.1500\linewidth]{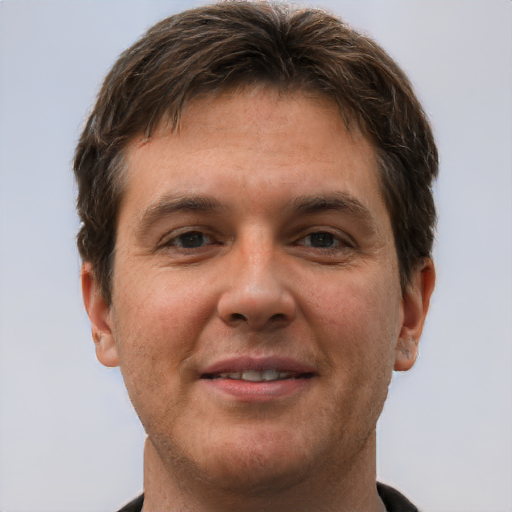} &
        \includegraphics[width=0.1500\linewidth]{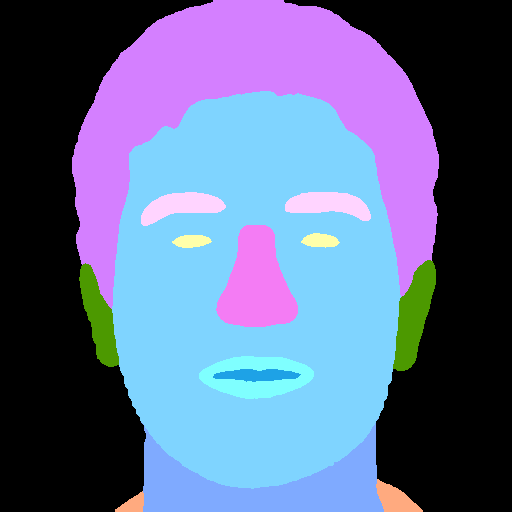} &
        \includegraphics[width=0.1500\linewidth]{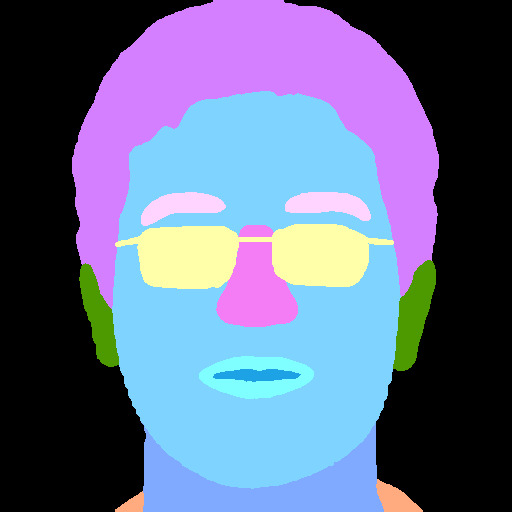} &
        \includegraphics[width=0.1500\linewidth]{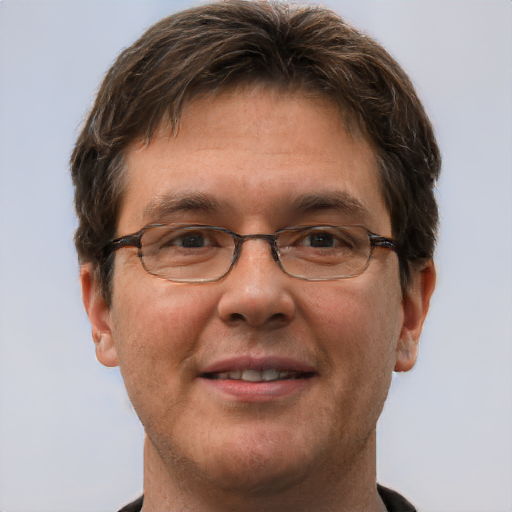} &
        \includegraphics[width=0.1500\linewidth]{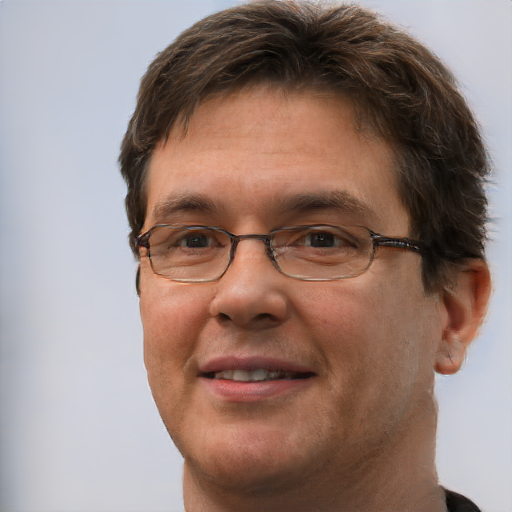} &
        \includegraphics[width=0.1500\linewidth]{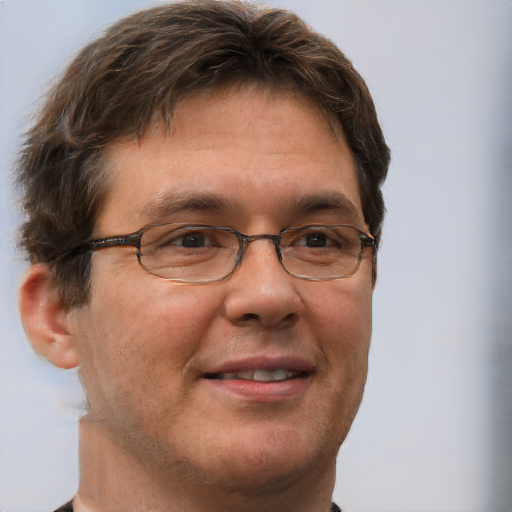}
        \\
        \rotatebox{90}{\hspace{2mm}+Nose} &
        \includegraphics[width=0.1500\linewidth]{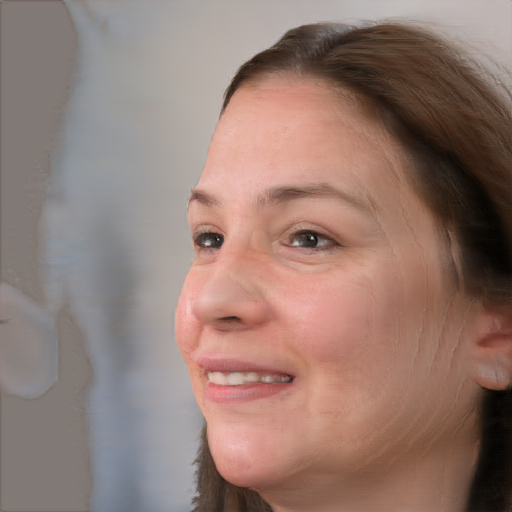} &
        \includegraphics[width=0.1500\linewidth]{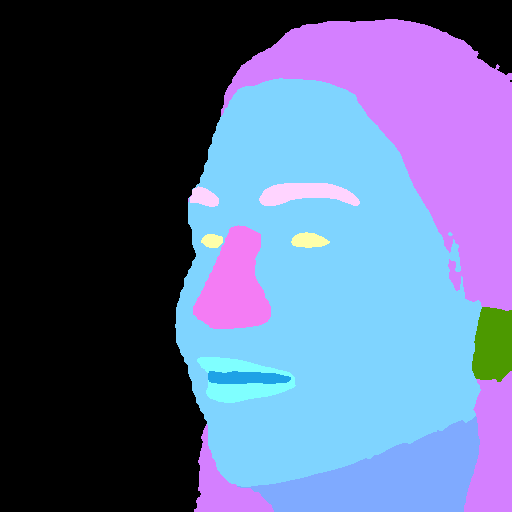} &
        \includegraphics[width=0.1500\linewidth]{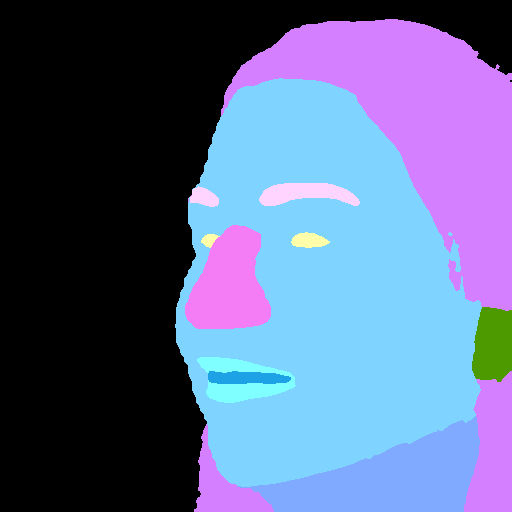} &
        \includegraphics[width=0.1500\linewidth]{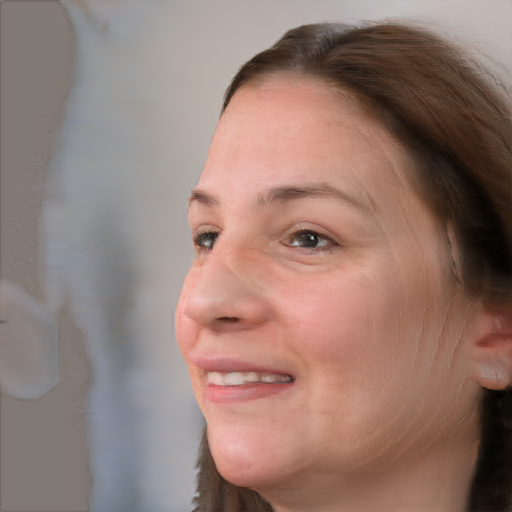} &
        \includegraphics[width=0.1500\linewidth]{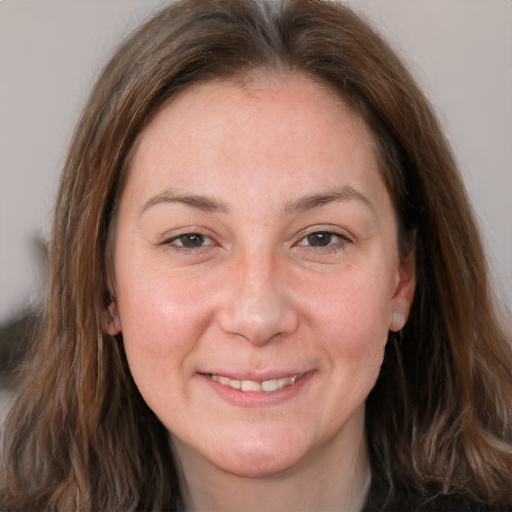} &
        \includegraphics[width=0.1500\linewidth]{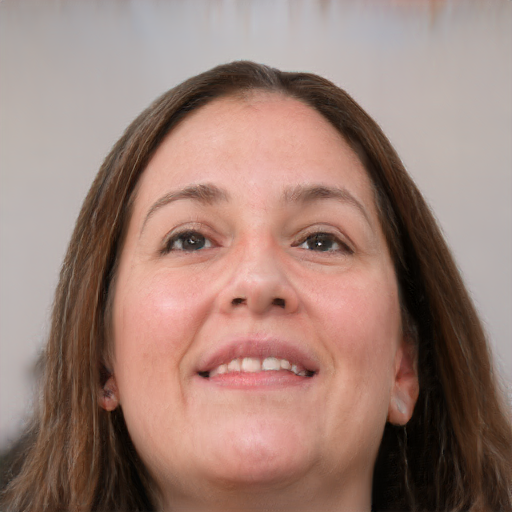}
        \\
        &
        \includegraphics[width=0.1500\linewidth]{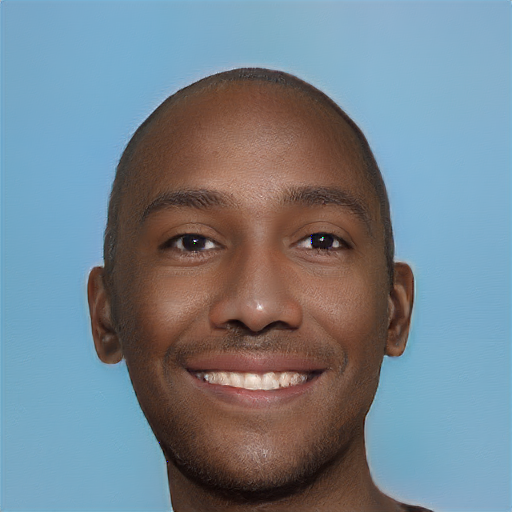} &
        \includegraphics[width=0.1500\linewidth]{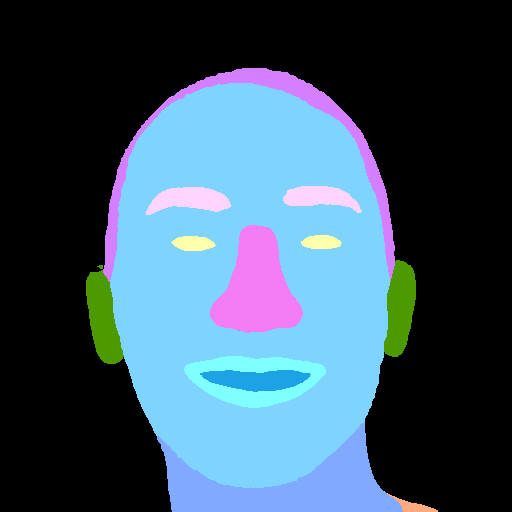} &
        \includegraphics[width=0.1500\linewidth]{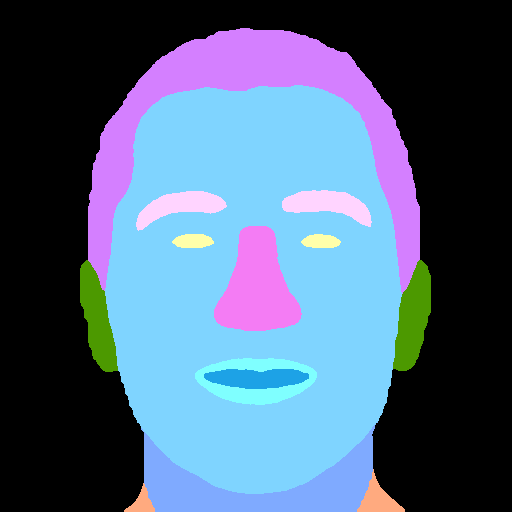} &
        \includegraphics[width=0.1500\linewidth]{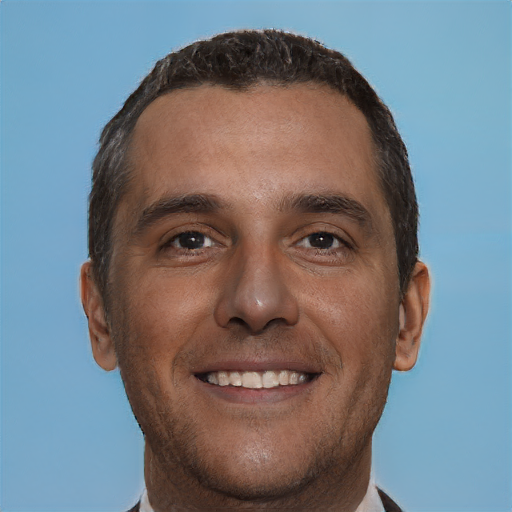} &
        \includegraphics[width=0.1500\linewidth]{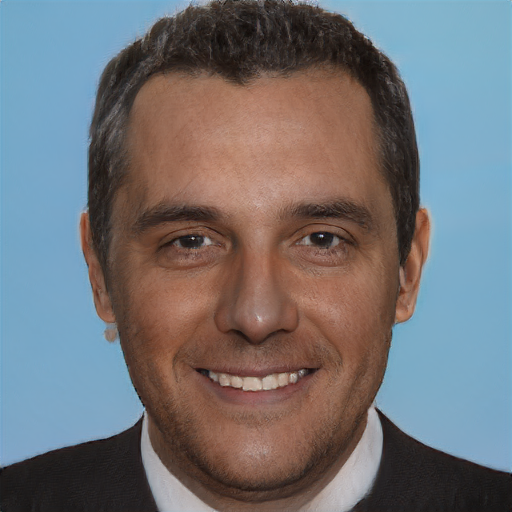} &
        \includegraphics[width=0.1500\linewidth]{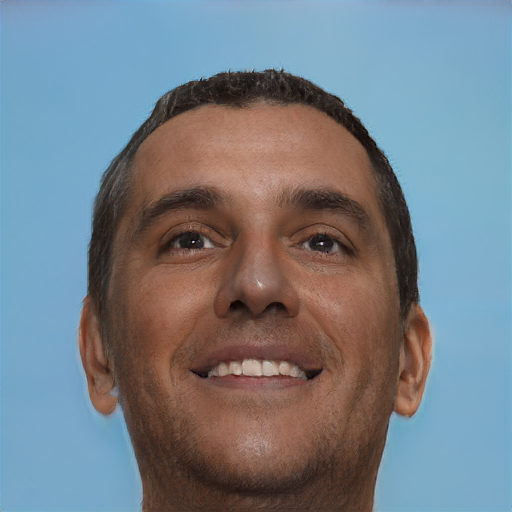}
        \\
        &
        \includegraphics[width=0.1500\linewidth]{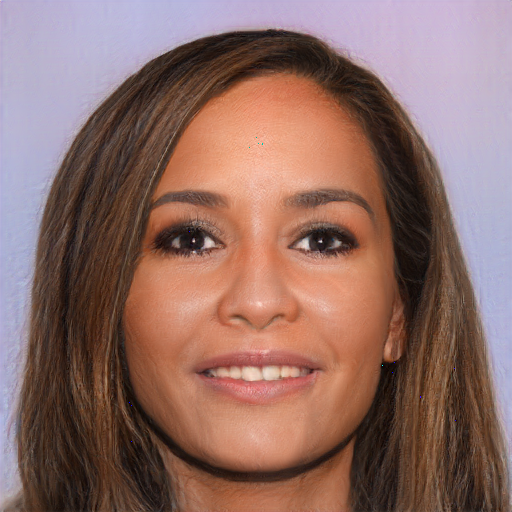} &
        \includegraphics[width=0.1500\linewidth]{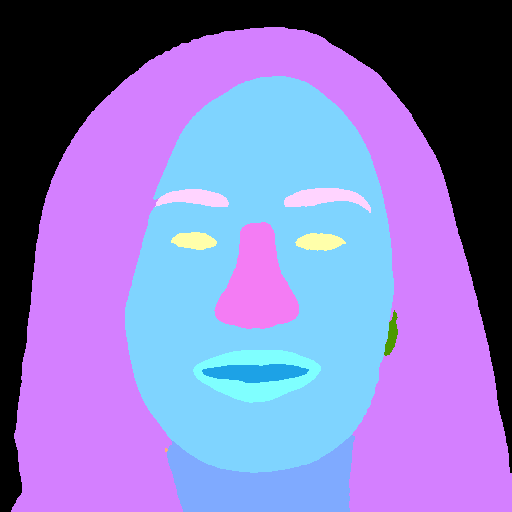} &
        \includegraphics[width=0.1500\linewidth]{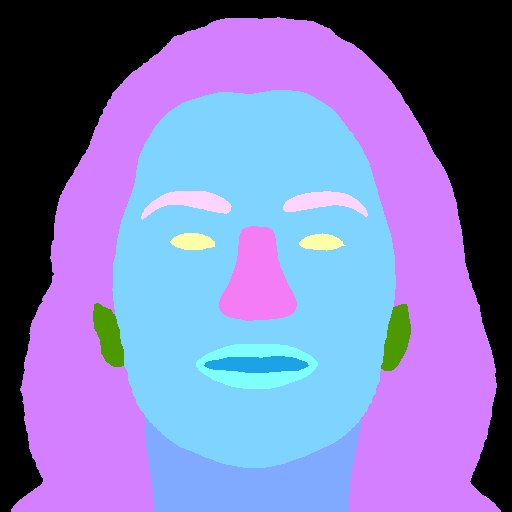} &
        \includegraphics[width=0.1500\linewidth]{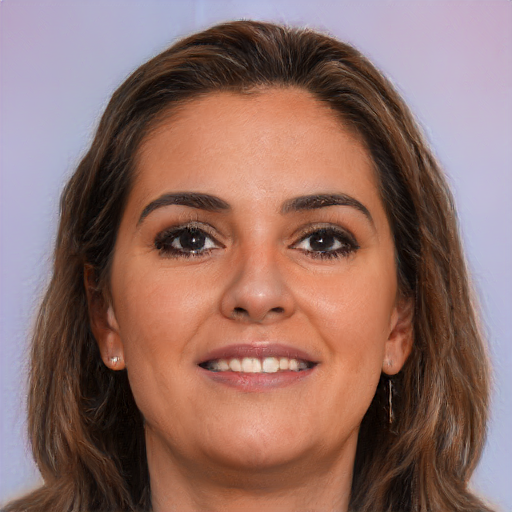} &
        \includegraphics[width=0.1500\linewidth]{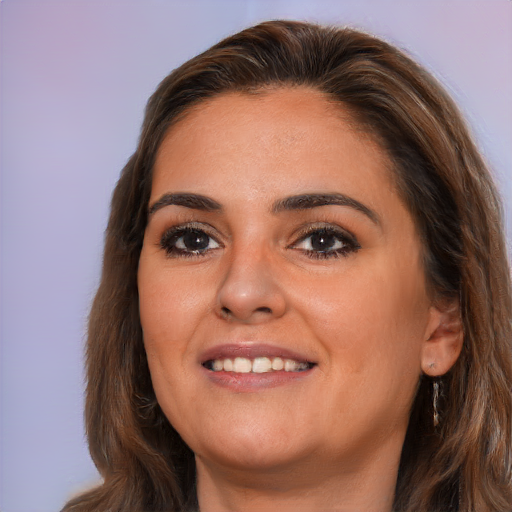} &
        \includegraphics[width=0.1500\linewidth]{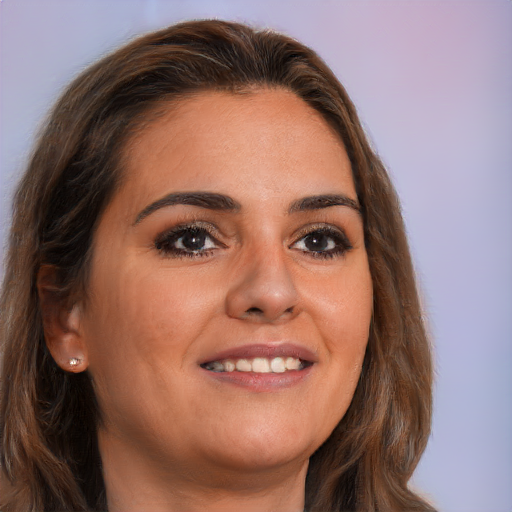}
    \end{tabular}
    
    \caption{{One- and multi-label} 
    editing. Our method enables {effective and intuitive} %
    {editing} guided by semantic masks at {frontal- and side-views}. %
    We manipulate facial attributes on the semantic map and use the optimization process described in Sec. \ref{editing-during-inference} {to obtain} the modified free-view portraits.}
    \label{fig:editing}
\end{figure}
\subsubsection*{Part-based Histogram Loss.}
To enhance disentanglement explicitly, we require rendered images to be similar in terms of color distribution including the background when 
{the CAM is fed with the same appearance features $\widehat{\mathbf{F}}_{\text{app}}$ but different normalized tri-planes}. %
For this purpose, we utilize the histogram loss introduced in \cite{afifi2021histogan}, which measures the similarity {among} 
histograms representing color distributions. However, we observe {that} its generated results generally transfer the style but fail to capture the style in detail, especially for hairs
(see the "Baseline3" of part (b) in Fig. \ref{fig:ablation_study}). Thus, we enhance this histogram loss by performing it for each label separately guided by semantic masks {(as illustrated in Fig. \ref{fig:promoting-similarity})}. 

Specifically, we sample a batch of latent codes $w(w^{(1)}, w^{(2)}, ..., w^{(B)})$ and apply the appearance feature $\widehat{\mathbf{F}}_{\text{app}}$ of one of the latent codes $w^{(k)}$ to the normalized tri-planes generated by other latent codes for the purpose of same appearance but different geometry. Assuming that there are \(N\) classes, {we define} the training procedure as: %
\hspace{-4mm}
\begin{multline}
    (\widehat{I}_{RGB}, \widehat{d}, \widehat{S})=\widetilde{G}(w, \widehat{\mathbf{F}}_{\text{app}}) \\
    \mathcal{L}_{Sim}=\lambda_5 \sum_{i=1}^{N}w_i\sum_{j=1}^{B}\mathcal{L}_{h}(\mathbf{H}(\hat{I}^{(j)}_{RGB} \odot M_i^{(j)}), \mathbf{H}(\hat{I}_{RGB}^{(k)} \odot M_i^{(k)})),
\end{multline}
where \(\sum_{i=1}^{N}w_i=1\), \(\mathbf{H}(\cdot)\) denotes the extraction of histogram features from images, \(\mathcal{L}_{h}\) stands for the distance {among}
histogram features, \(\hat{I}_{RGB}\) represents rendered images from %
\(\widehat{F}_{\text{app}}\), \(M_i\) denotes the mask for label \(i\) in the semantic masks \(\widehat{S}\) corresponding to \(\hat{I}_{RGB}\), and \(B\) represents the batch size. In our experiments, we empirically set \(\lambda_5=15, B=3, k=1\).

Our final objective \(\mathcal{L}\) is simply the sum of the above two losses (as we have already weighted each term in these losses){: $    \mathcal{L} = \mathcal{L}_{Recon}+\mathcal{L}_{Sim}$. Minimizing} \(\mathcal{L}\) will lead to the optimization of three networks: \(\Psi_{\text{geo}}, \Psi_{\text{app}}, \widetilde{f}\).

\subsection{Editing during Inference}
\label{editing-during-inference}
We can edit samples \(\mathbf{I}\) at {a certain} pose generated by {a} latent code \(w\in W\) space or real images.
{To edit given} real face images \(\mathbf{I'}\), we first invert the images into the \(\mathcal{W}\) space as \(w\) with pivotal tuning inversion \cite{roich2021pivotal} following EG3D{,} denoted as \(\mathbf{I}\).
In all {the editing} cases, a user-edited semantic mask \(\hat{S}\) is assumed to be available given {an} original semantic mask \(S\).

Formally, we are seeking an editing vector \(\delta w^{+} \in \mathcal{W}^{+}\) such that \((\mathbf{I}_{\text{edited}}, \mathbf{S}_{\text{edited}})=G'(w+\delta w^{+}, \mathbf{F}_{\text{app}}(w))\), in which $\mathbf{I}_{\text{edited}}$ is generated by the optimized latent code and \(\mathbf{S}_{\text{edited}}\) approximates \(\hat{\mathbf{S}}\). Note that the appearance feature is kept fixed
during optimization to keep {the} appearance unchanged.
As in EditGAN \cite{ling2021editgan}, we first define {a} region of interest \(r\) within which we expect the image to change due to {a certain edit}. %
{A sequence of such {edits} can be achieved step by step and is illustrated in the supplementary.}

To optimize \(\delta w^{+}\) so that \(\mathbf{S}_{\text{edited}}\) approximates \(\hat{S}\) while preserving regions {outside of $r$} %
untouched, we use the following losses as the minimization targets:
\begin{multline}
    \mathcal{L}_{\text{editing}}(\delta w^{+})= \mathcal{L}_{\text{VGG}}(\mathbf{I}_{\text{edited}} \odot (1-r), \mathbf{I} \odot (1-r)) \\
    + \mathcal{L}_{\text{MSE}}(\mathbf{I}_{\text{edited}} \odot (1-r), \mathbf{I} \odot (1-r)) + E(\mathbf{S}_{\text{edited}}, \hat{\mathbf{S}}),
\end{multline}
{where $\mathcal{L}_{\text{MSE}}$ denotes the mean square error.}

The only "learnable" variable is the editing vector \(\delta w^{+}\) and all {the} neural networks are kept fixed. Note that as in EditGAN, there is a certain amount of {ambiguity} in how the segmentation modification is realized in the RGB output.

\section{Experiments}
In this section, we show our experimental setup {and} discuss the results of our experiments. {Results} from comparison with alternative methods, ablation study, and user study {all} show the effectiveness of our method
and its superiority to the alternative approaches.

\subsubsection*{Implementation Details.} Our decoder is {implemented} as a lightweight MLP with one hidden layer of 64 units. We use the Adam \cite{kingma2014adam} optimizer with \(\beta_1=0,\beta_2=0.99\) and the learning rate is fixed to \(0.002\) during fine-tuning. The fine-tuning takes 96 hours on 1 Tesla V100 GPU. NeRFFaceEditing is implemented in PyTorch \cite{PyTorch} and Jittor \cite{hu2020jittor}.

\subsection{Results and Evaluations}
In terms of rendering speed at inference {time}, our method achieves nearly real-time framerates at $512^2$ resolution. 
{On a single Tesla V100 GPU, we could render $20$ images per second without tri-plane caching or $32$ with tri-plane caching.}
{Besides, we} make comparisons {by inverting real images in the wild} with {SofGAN \cite{chen2022sofgan},} {which is also a%
} view-consistent method.
\subsubsection*{Qualitative results.}
\begin{figure}
    \renewcommand\tabcolsep{0.0pt}
    \renewcommand{\arraystretch}{0}
    \centering \small
    \begin{tabular}{ll}
         (a) &  
         \begin{minipage}[t]{0.50\textwidth}
         \begin{tabular}{cccc|cc|cc}
            \rotatebox{90}{\hspace{1mm}-Mouth}
            & \includegraphics[width=0.1285\linewidth]{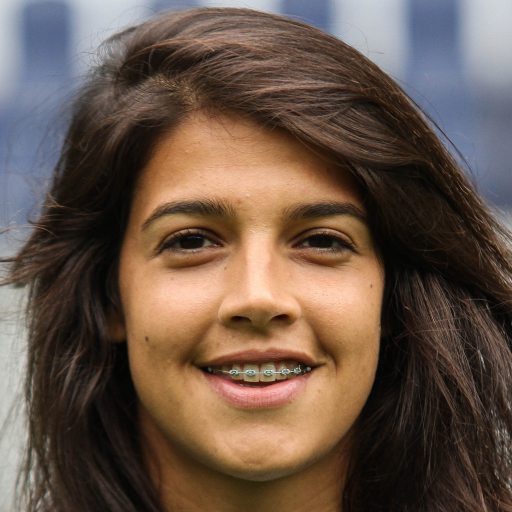}
            & \includegraphics[width=0.1285\linewidth]{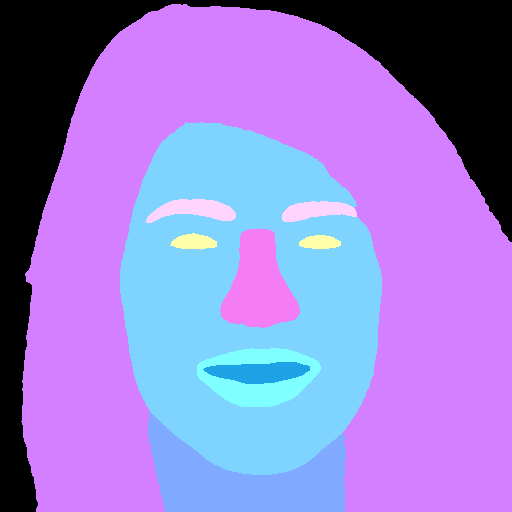}
            & \includegraphics[width=0.1285\linewidth]{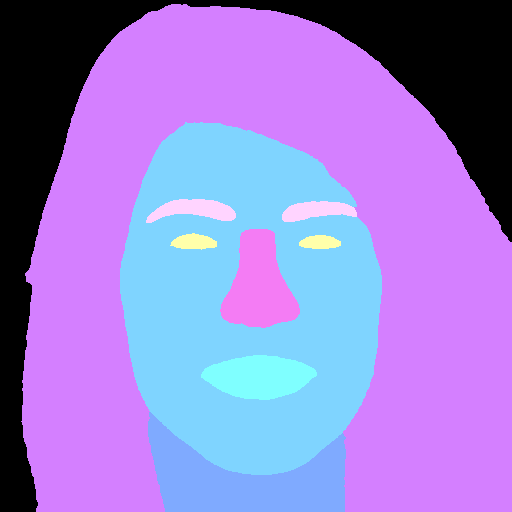}
            & \includegraphics[width=0.1285\linewidth]{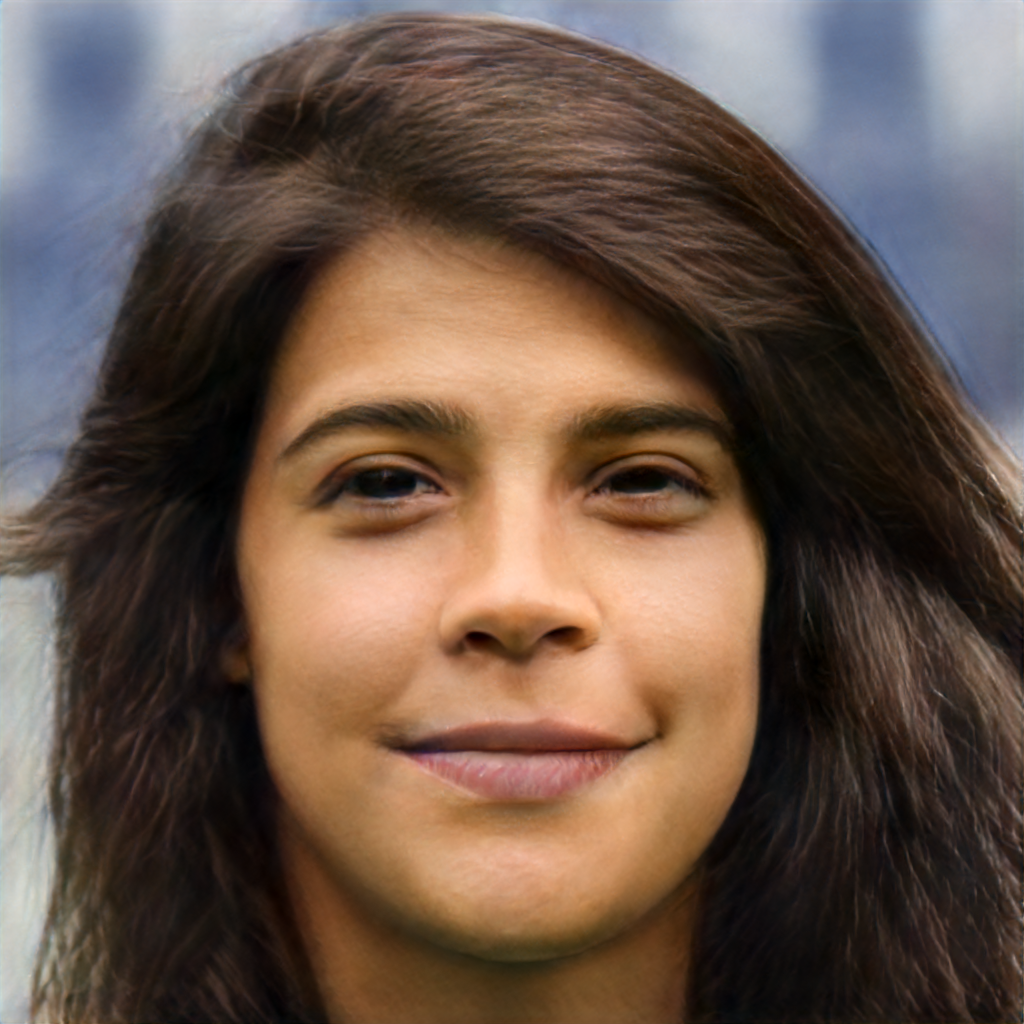}
            & \includegraphics[width=0.1285\linewidth]{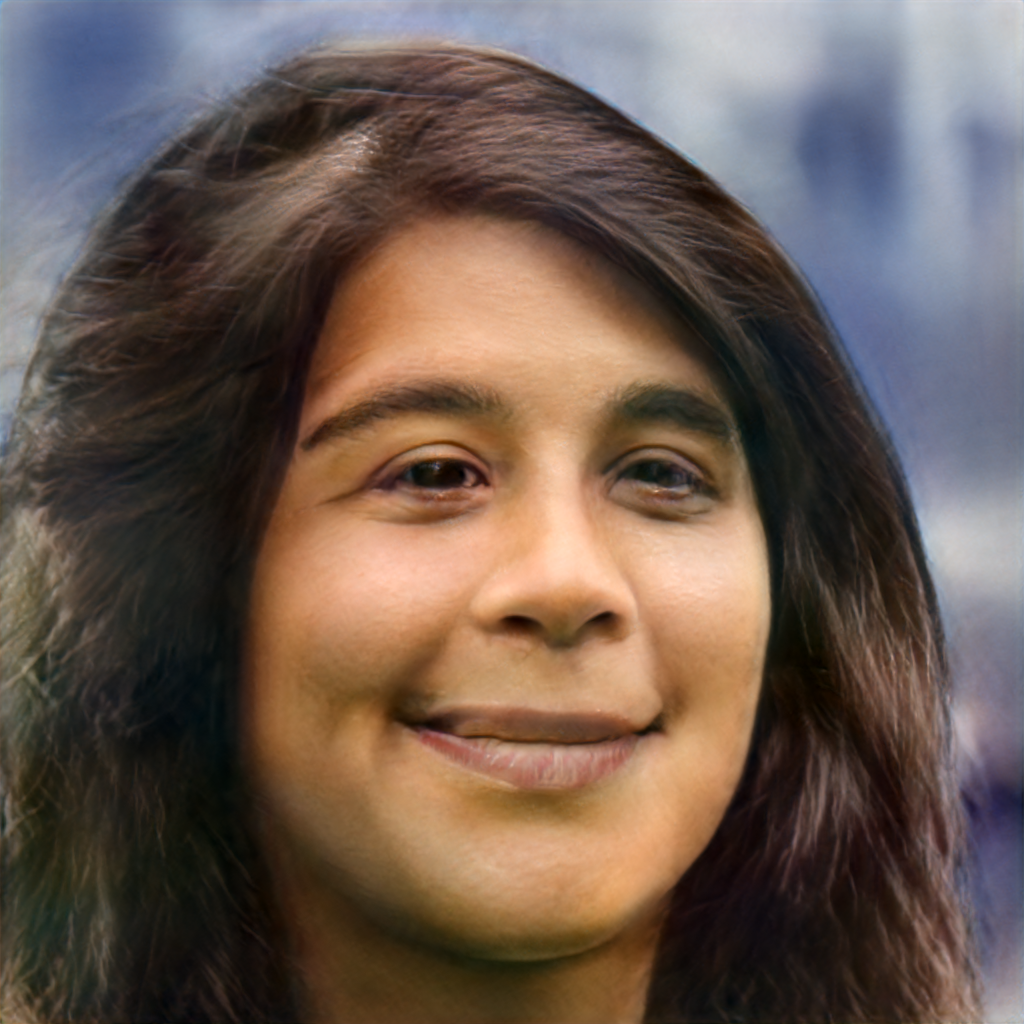}
            & \includegraphics[width=0.1285\linewidth]{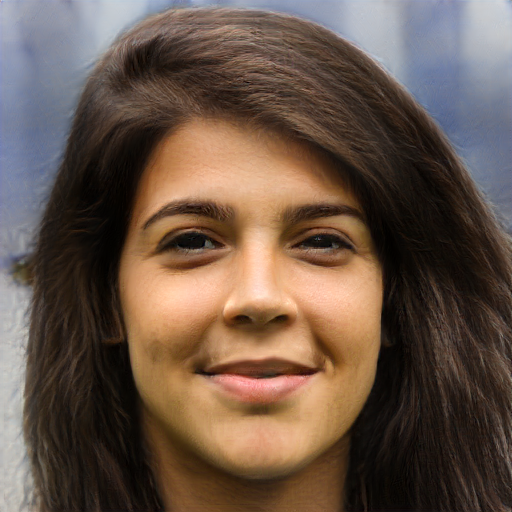}
            & \includegraphics[width=0.1285\linewidth]{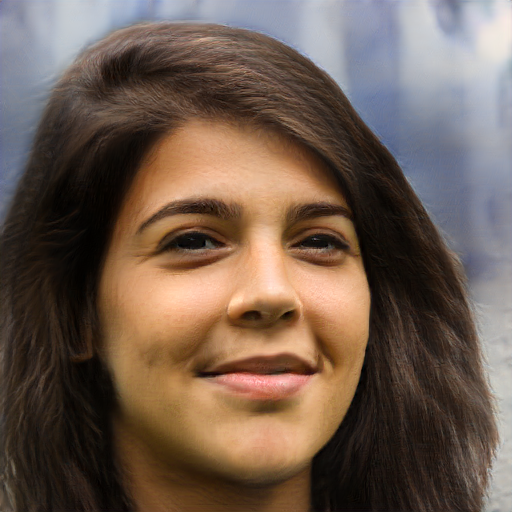}
            \\
            \rotatebox{90}{\hspace{1mm}+Skin}
            & \includegraphics[width=0.1285\linewidth]{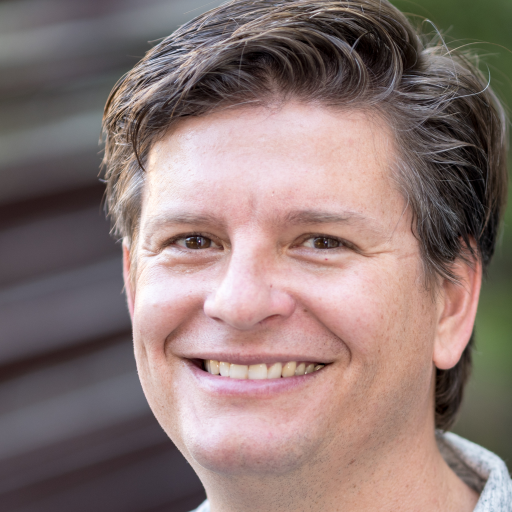}
            & \includegraphics[width=0.1285\linewidth]{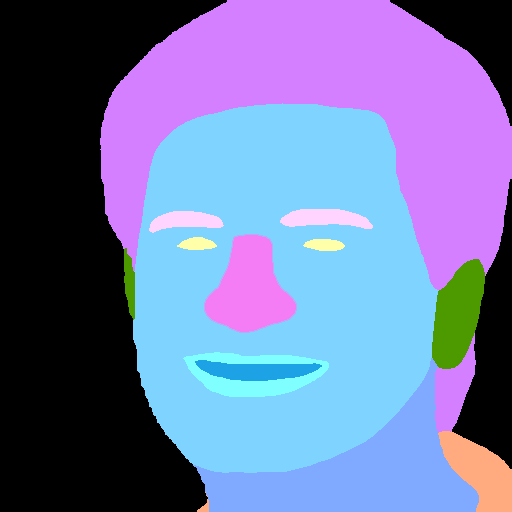}
            & \includegraphics[width=0.1285\linewidth]{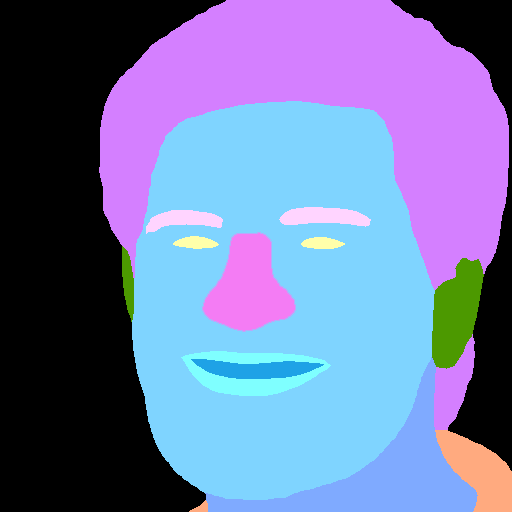}
            & \includegraphics[width=0.1285\linewidth]{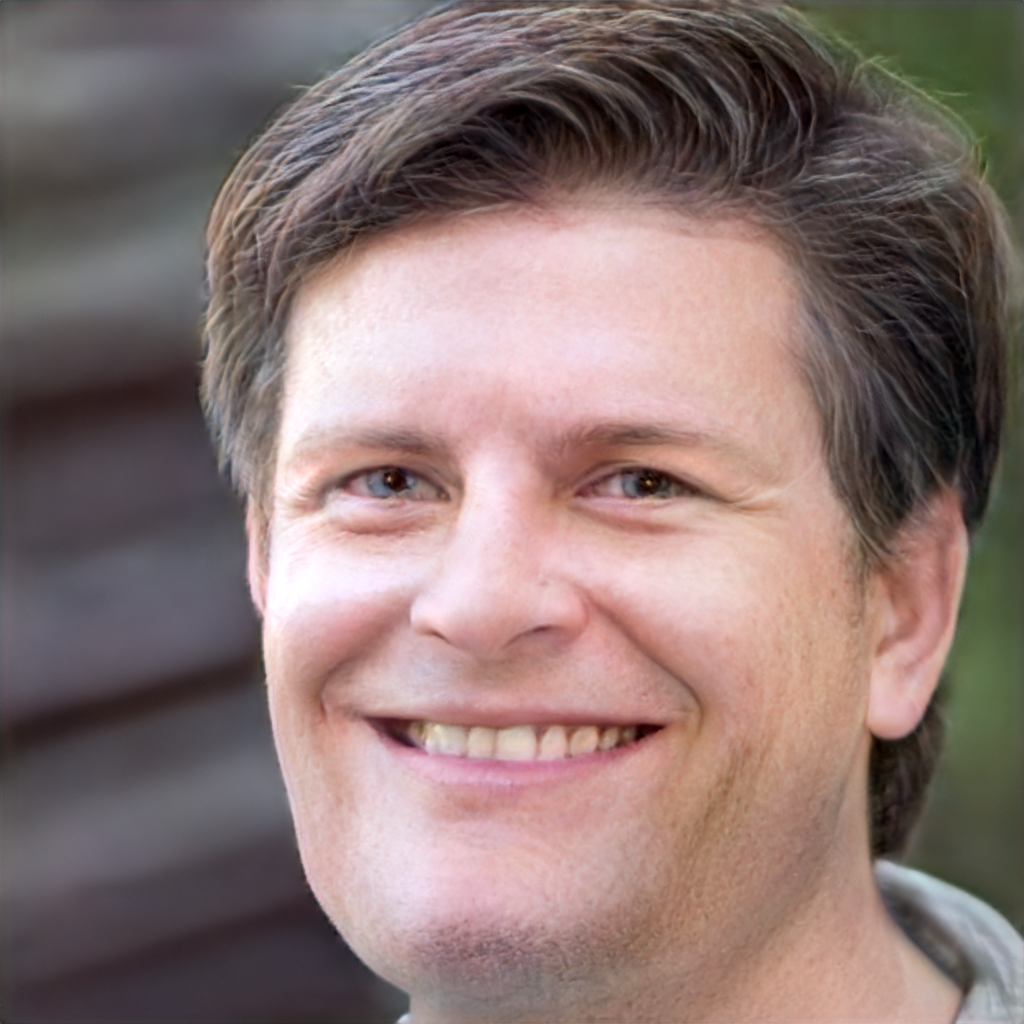}
            & \includegraphics[width=0.1285\linewidth]{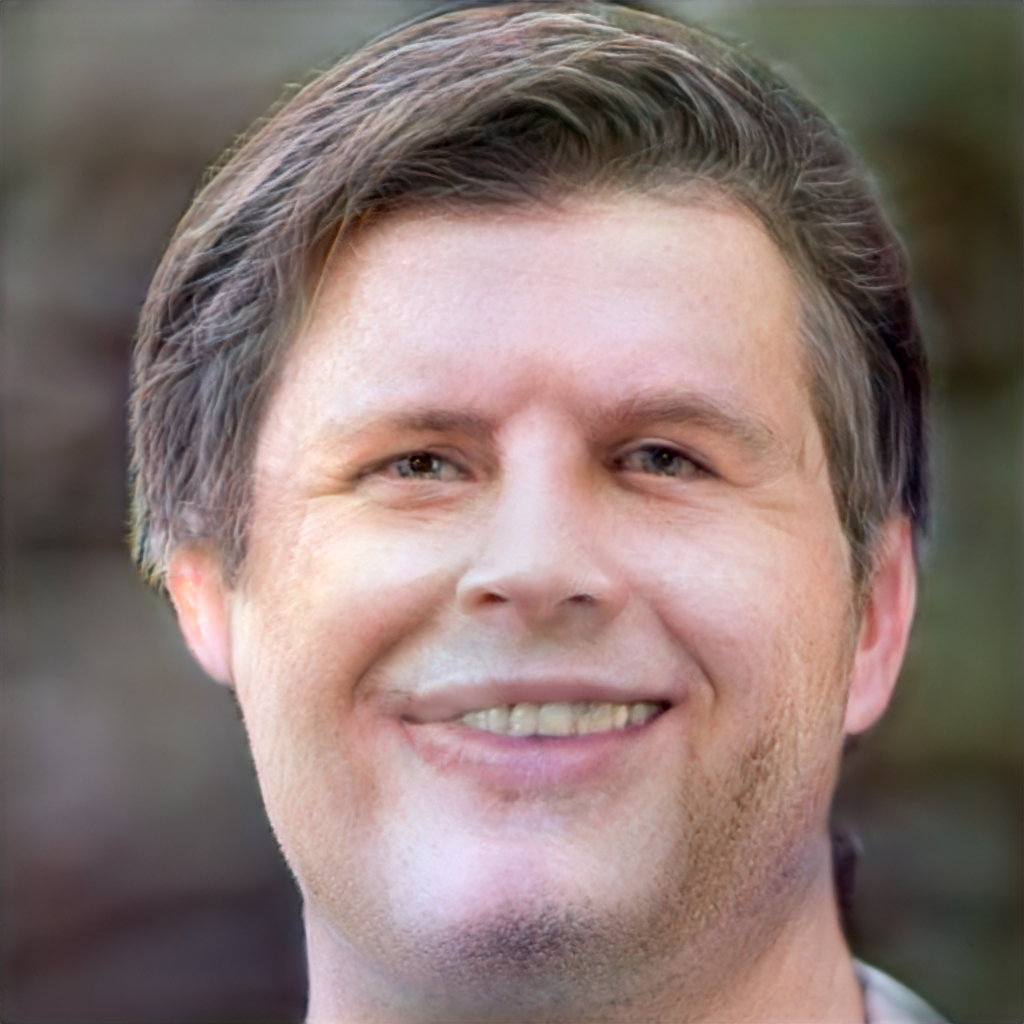}
            & \includegraphics[width=0.1285\linewidth]{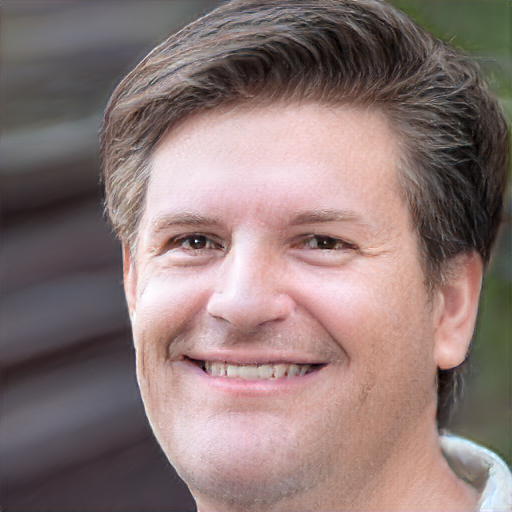}
            & \includegraphics[width=0.1285\linewidth]{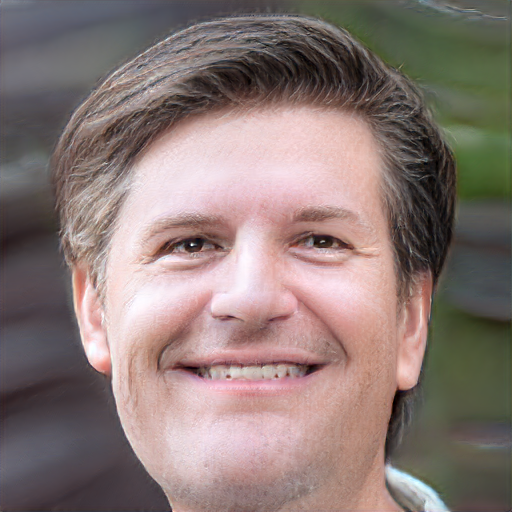}
            \\
             & Source & Mask & Target & \multicolumn{2}{c}{(a) {SofGAN}} &  \multicolumn{2}{c}{(b) {Ours}}
         \end{tabular}
         \end{minipage}
         \\
         \\
         \vspace{2mm}
         \\
         (b) &
         \begin{minipage}[t]{0.47\textwidth}
             \begin{tabular}{cc|cc|cc}
                \includegraphics[width=0.1666\linewidth]{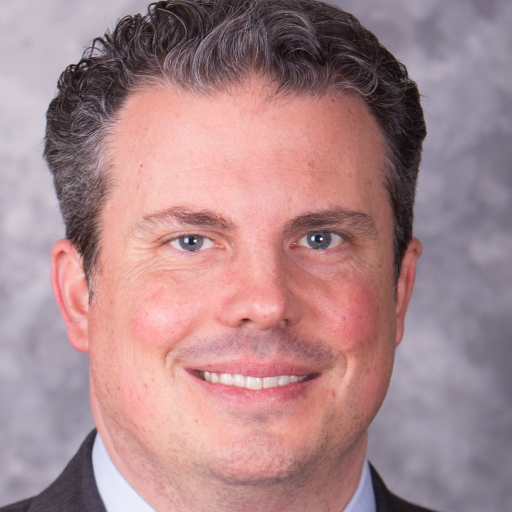}
                & \includegraphics[width=0.1666\linewidth]{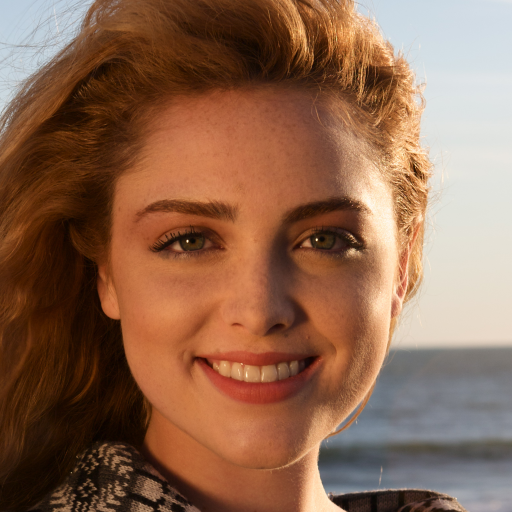}
                & \includegraphics[width=0.1666\linewidth]{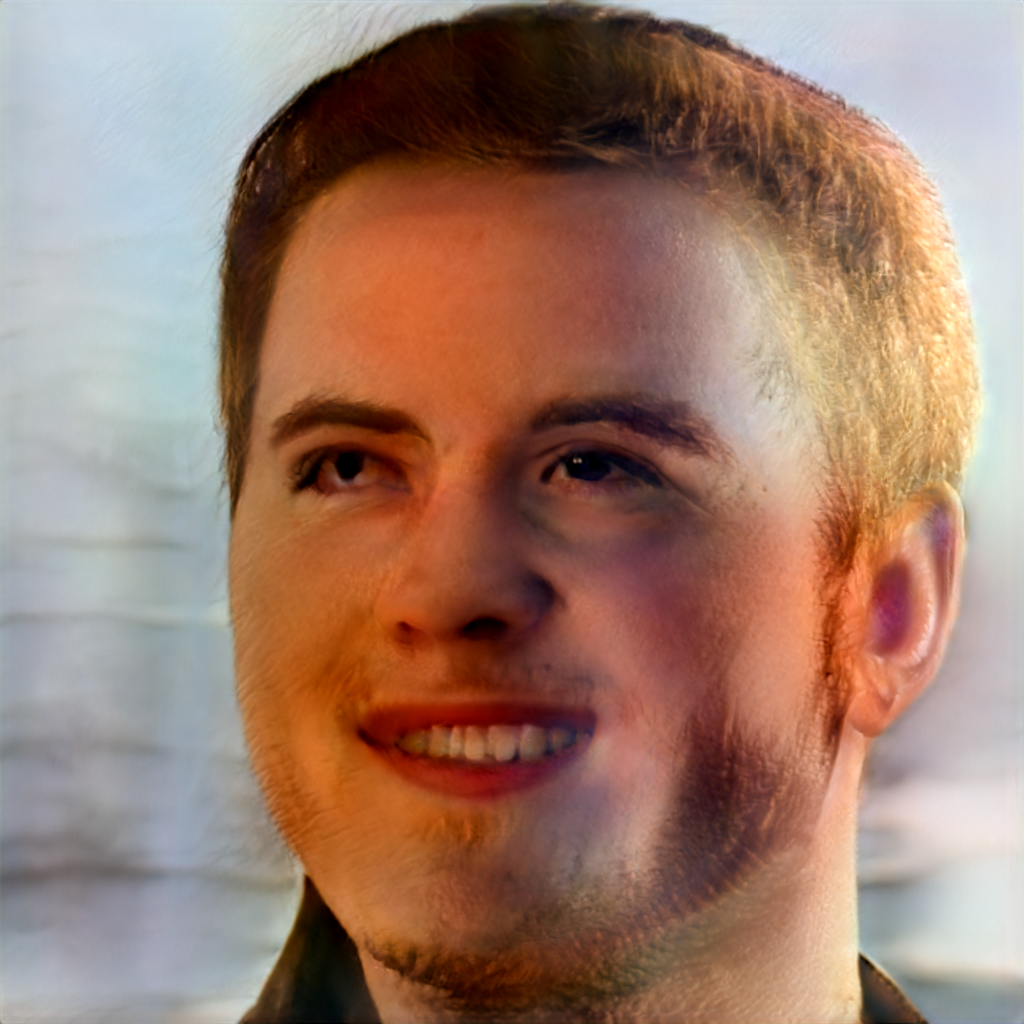}
                & \includegraphics[width=0.1666\linewidth]{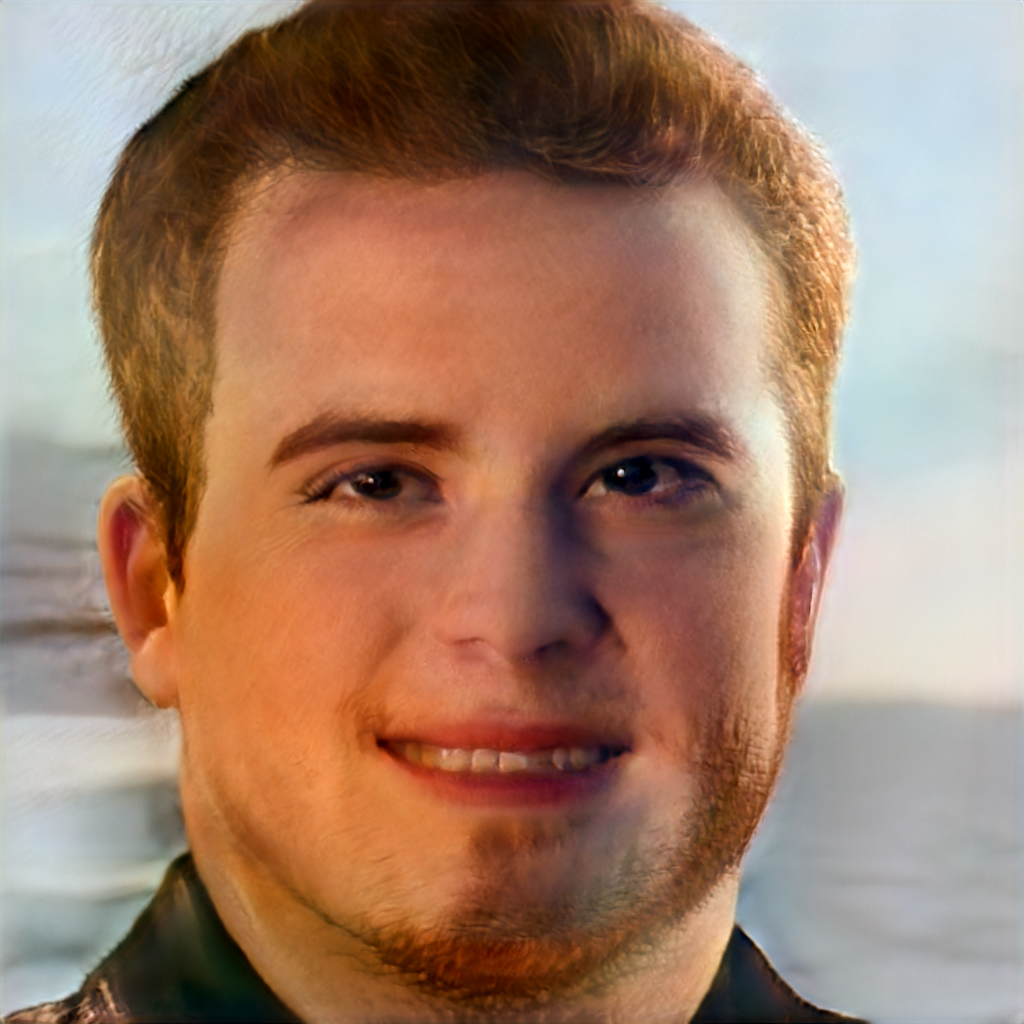}
                & \includegraphics[width=0.1666\linewidth]{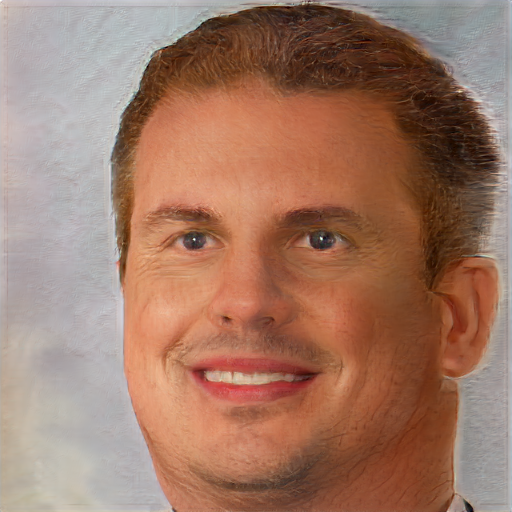}
                & \includegraphics[width=0.1666\linewidth]{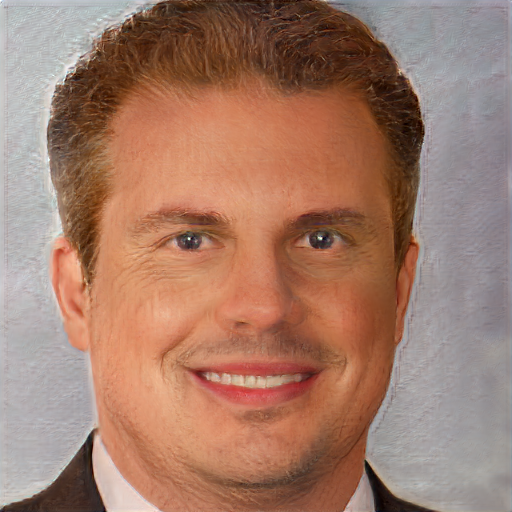}
                \\
                \includegraphics[width=0.1666\linewidth]{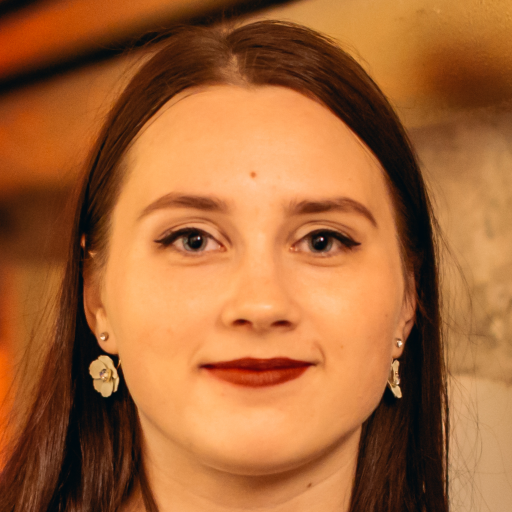}
                & \includegraphics[width=0.1666\linewidth]{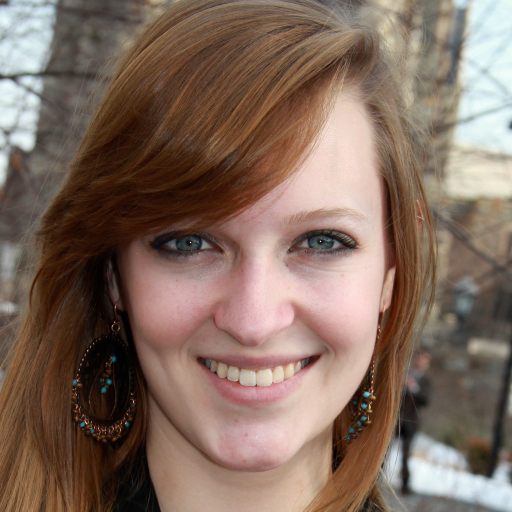}
                & \includegraphics[width=0.1666\linewidth]{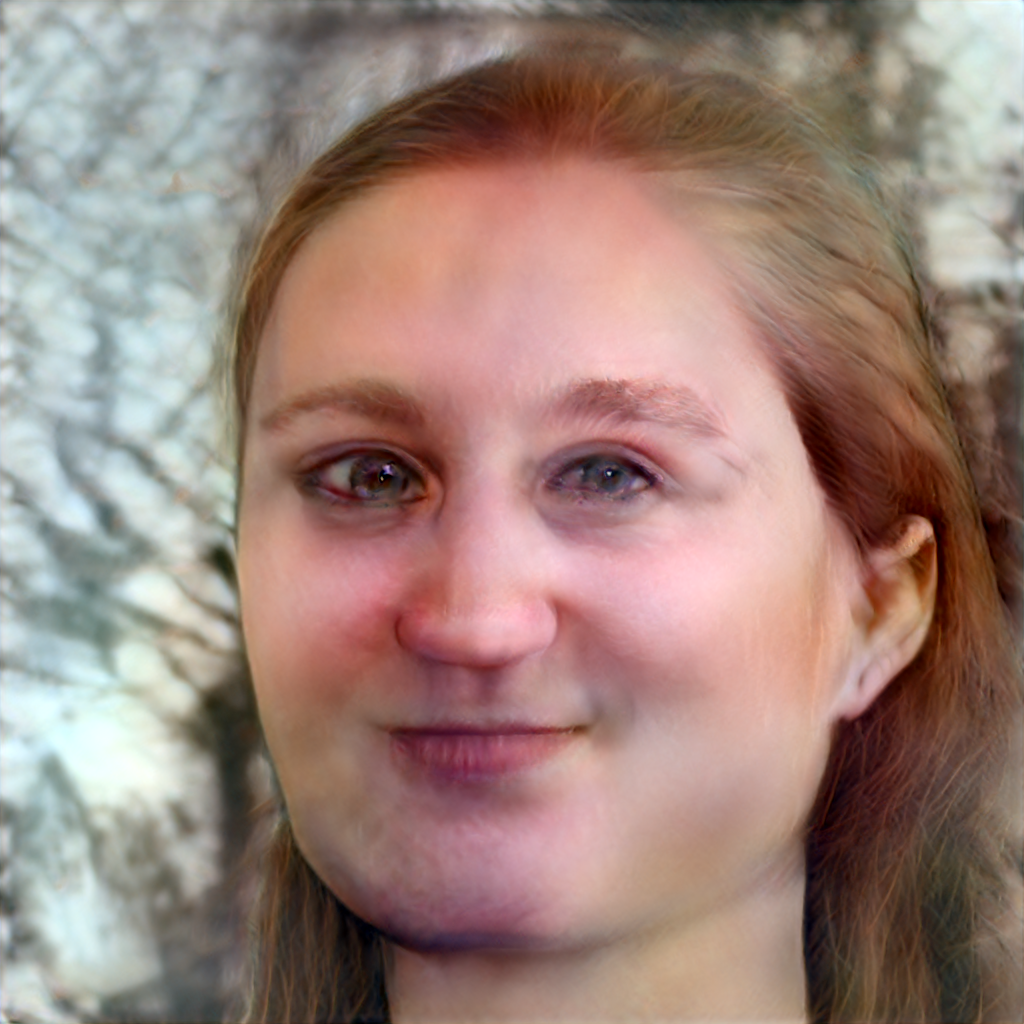}
                & \includegraphics[width=0.1666\linewidth]{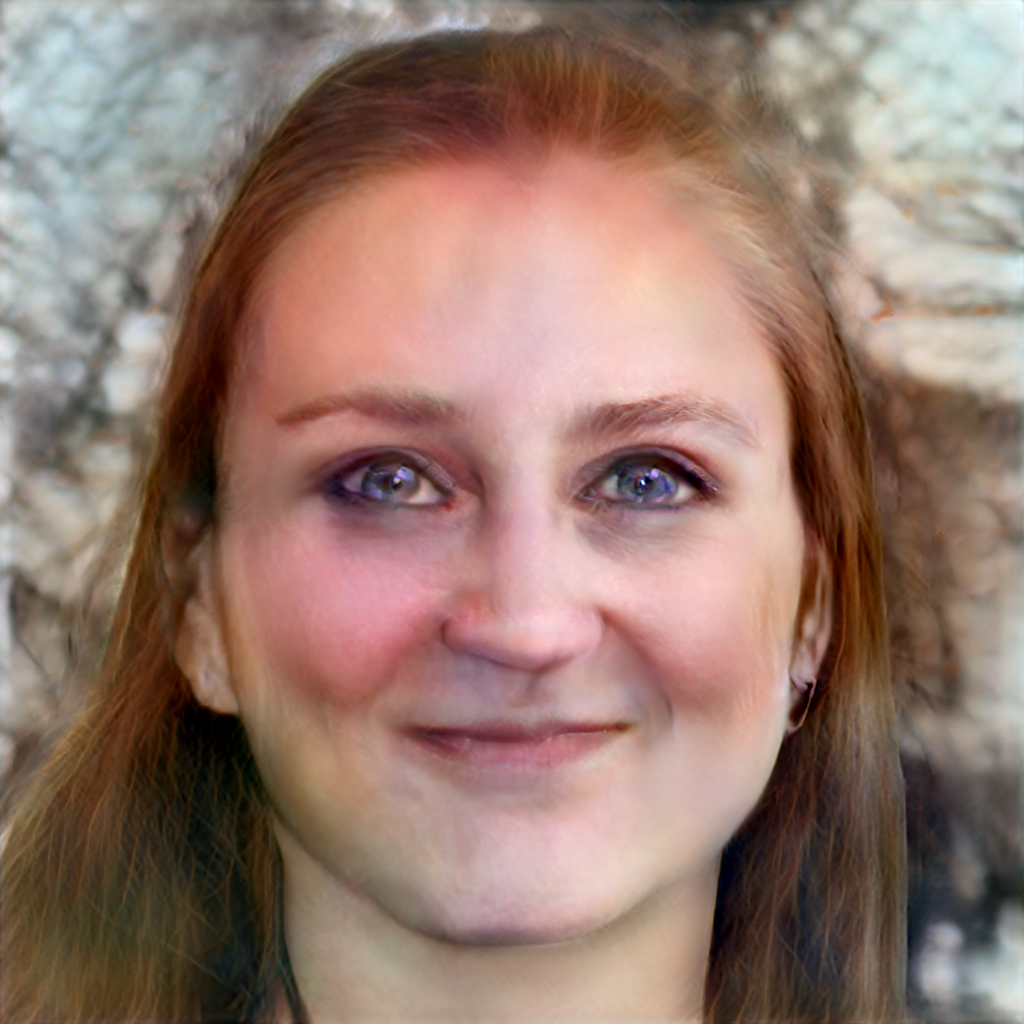}
                & \includegraphics[width=0.1666\linewidth]{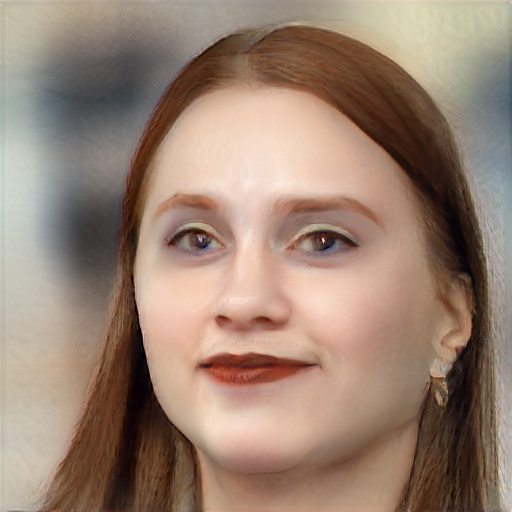}
                & \includegraphics[width=0.1666\linewidth]{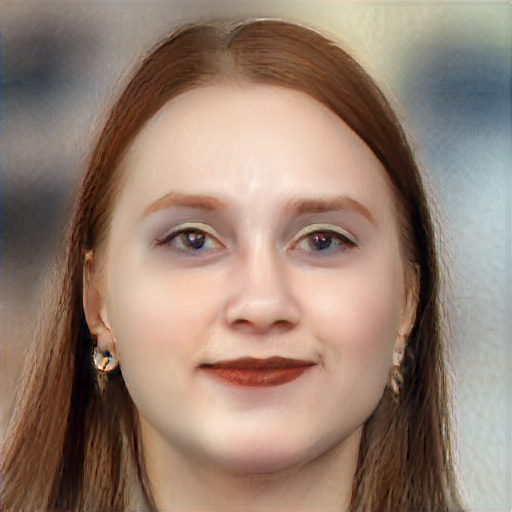}
                \\
                Geo. & App. & \multicolumn{2}{c}{(a) SofGAN Free-viewed} & \multicolumn{2}{c}{(b) Ours Free-viewed}
            \end{tabular}
         \end{minipage}
    \end{tabular}
    \caption{
    Comparisons with SofGAN \cite{chen2022sofgan} in terms of editing (a) and style transfer (b). Original images courtesy of Thomas Rodenbücher, Brett Morrison, United Way of Central Ohio, Colin Brown, Krists Luhaers and Mckinnon de Kuyper.
    }
    \label{fig:comparison}
\end{figure}

Our framework can generate semantic masks as well as realistic images from a certain pose. Thus, it can be used to edit {the} original facial volume through GAN inversion (described in Sec. \ref{editing-during-inference}), {as shown in Fig. \ref{fig:editing} (please see the supplementary materials for more results)}. {For comparison, we perform editing on the real images as shown in {Fig. \ref{fig:comparison}(a)}.} %
SofGAN achieves reasonable results at original poses, but disturbs the identity at other poses and fails to {produce facial details (e.g., in the mouth and eyes) naturally}.
In contrast, our method outperforms SofGAN
in both identity preservation and editing fidelity.
\begin{figure}
    \renewcommand\tabcolsep{0.0pt}
    \renewcommand{\arraystretch}{0}
    \centering \small
    \begin{tabular}{ll}
            (a) &  
            \begin{minipage}[t]{0.47\textwidth}
                \begin{tabular}{cccccccc}
            \rotatebox{90}{\hspace{2mm}-Hair}
            & \includegraphics[width=0.14\linewidth]{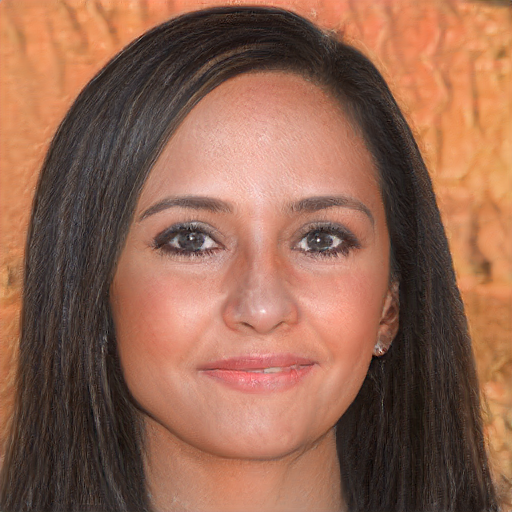}
            & \includegraphics[width=0.14\linewidth]{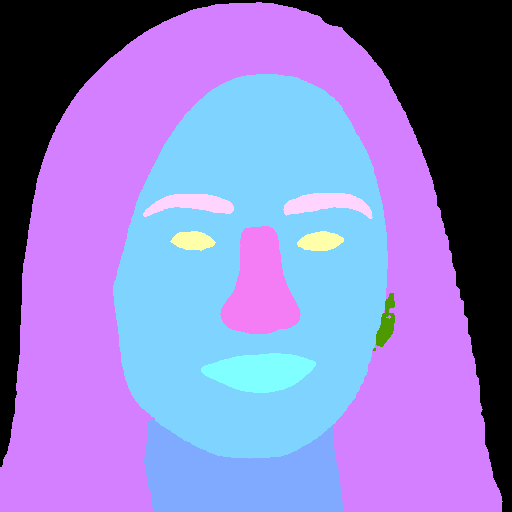}
            & \includegraphics[width=0.14\linewidth]{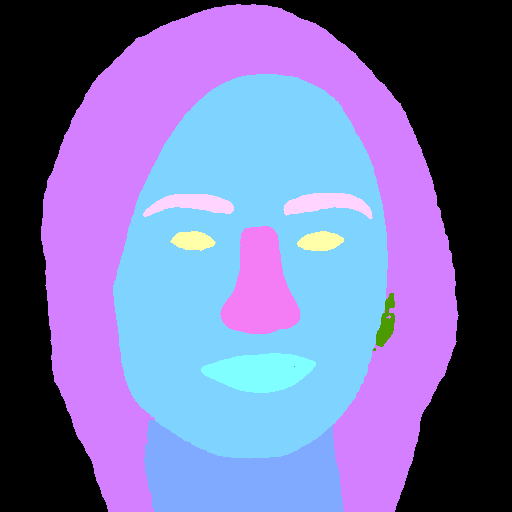}
            & \includegraphics[width=0.14\linewidth]{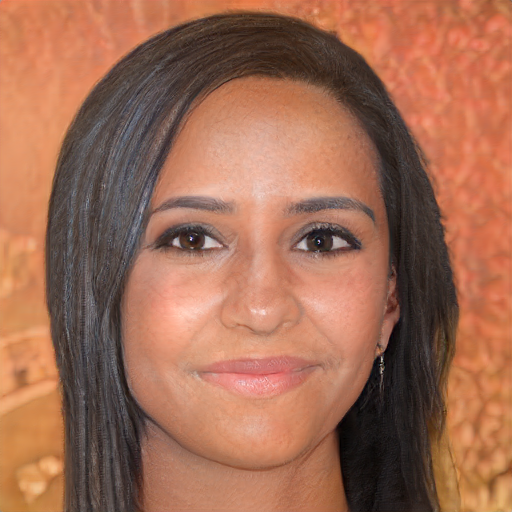}
            & \includegraphics[width=0.14\linewidth]{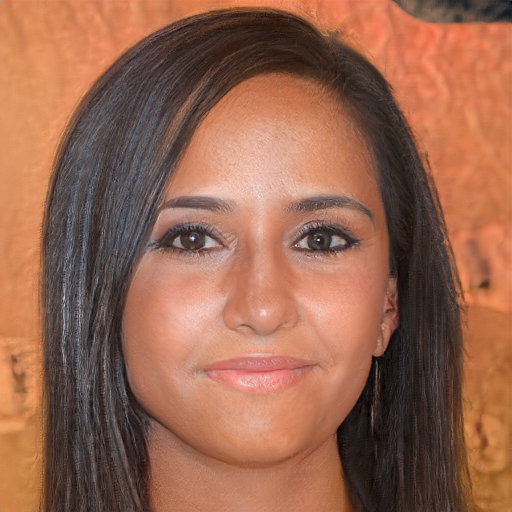}
            & \includegraphics[width=0.14\linewidth]{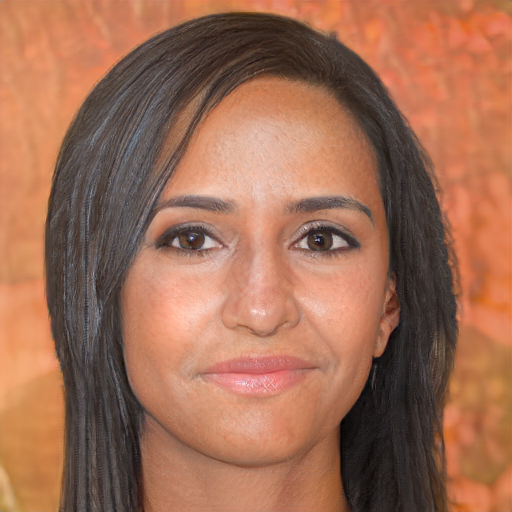}
            & \includegraphics[width=0.14\linewidth]{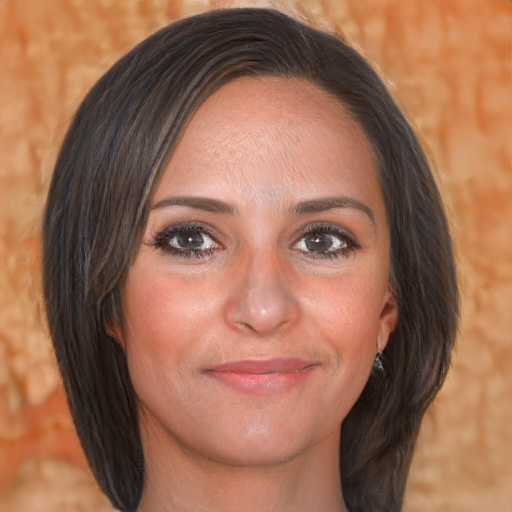}
            \\
            \rotatebox{90}{\hspace{1mm}+Glass}
            & \includegraphics[width=0.14\linewidth]{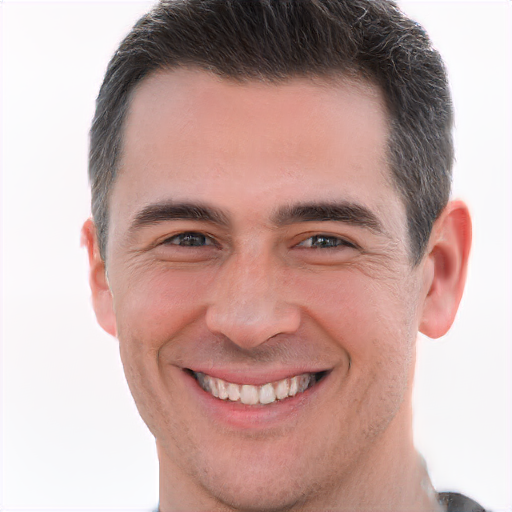}
            & \includegraphics[width=0.14\linewidth]{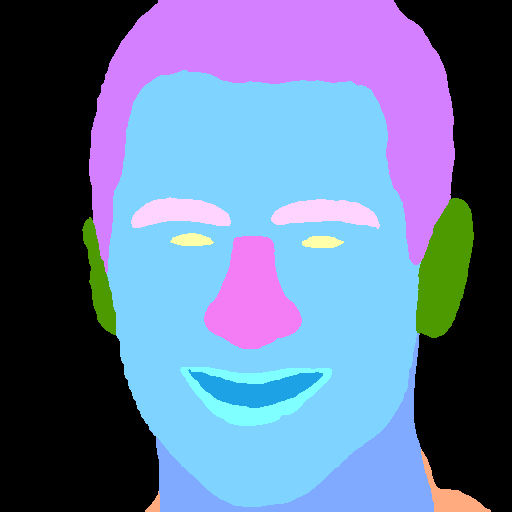}
            & \includegraphics[width=0.14\linewidth]{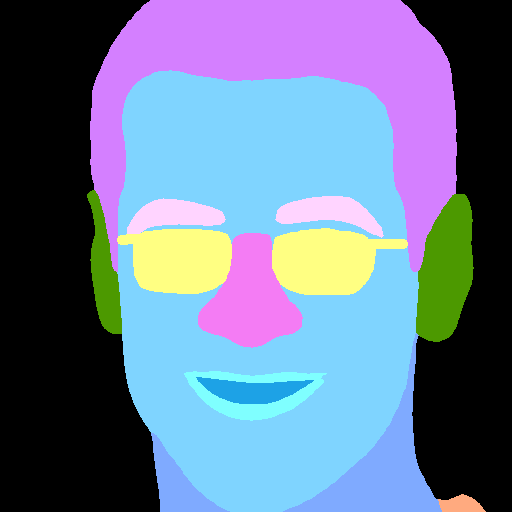}
            & \includegraphics[width=0.14\linewidth]{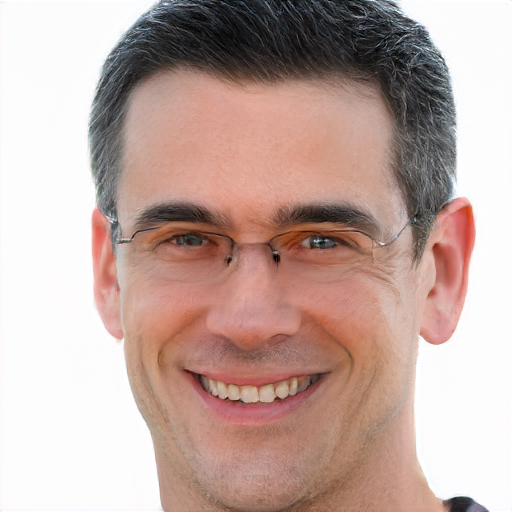}
            & \includegraphics[width=0.14\linewidth]{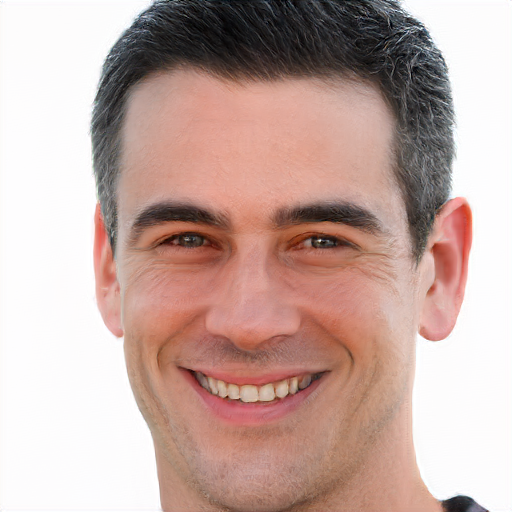}
            & \includegraphics[width=0.14\linewidth]{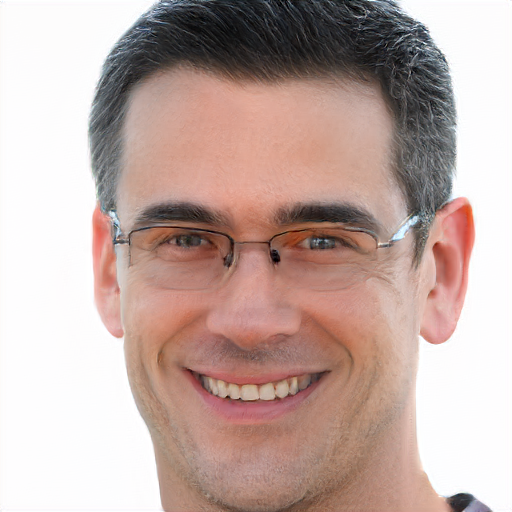}
            & \includegraphics[width=0.14\linewidth]{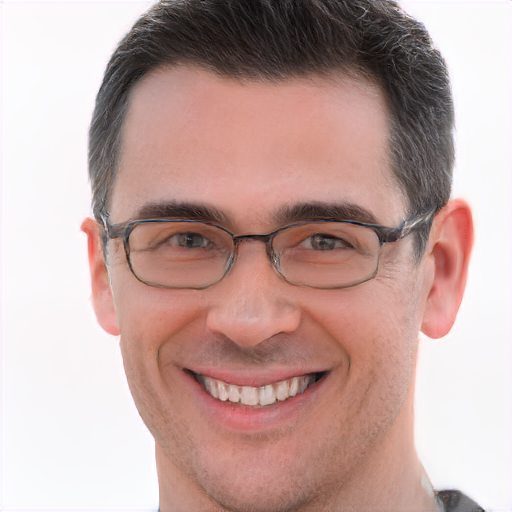}
            \\
             & Source & Mask & Modified & Baseline1 & Baseline2 & Baseline3 & Ours
        \end{tabular}
            \end{minipage}
            \\
            \\
            \vspace{2mm}
           \\
            (b) & 
            \begin{minipage}[t]{0.47\textwidth}
            \begin{tabular}{cccccc}
        \includegraphics[width=0.1666\linewidth]{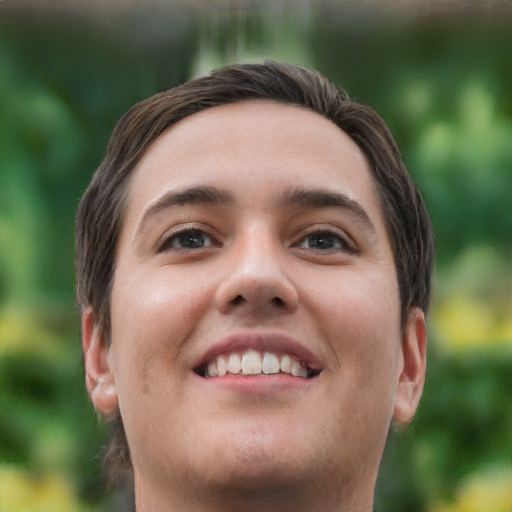}
        & \includegraphics[width=0.1666\linewidth]{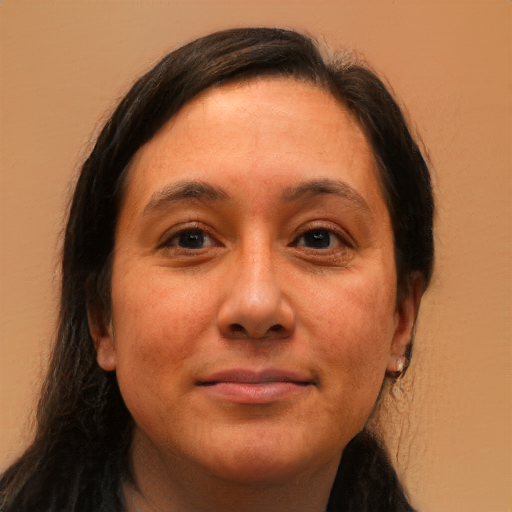}
        & \includegraphics[width=0.1666\linewidth]{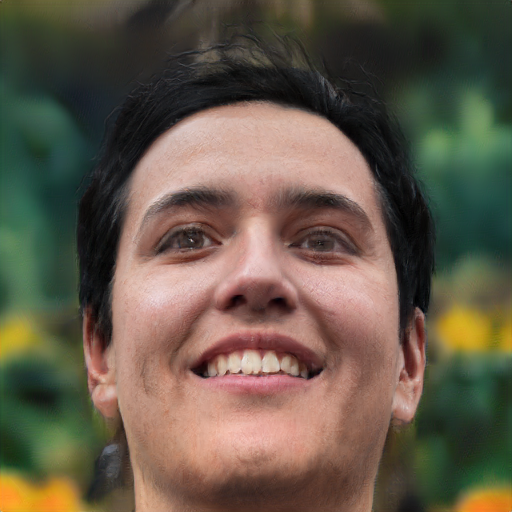}
        & \includegraphics[width=0.1666\linewidth]{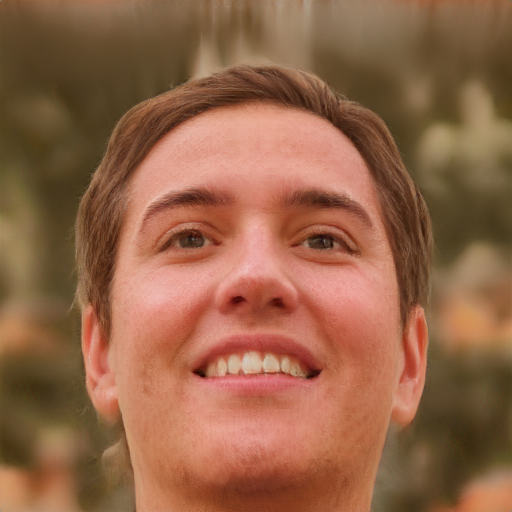}
        & \includegraphics[width=0.1666\linewidth]{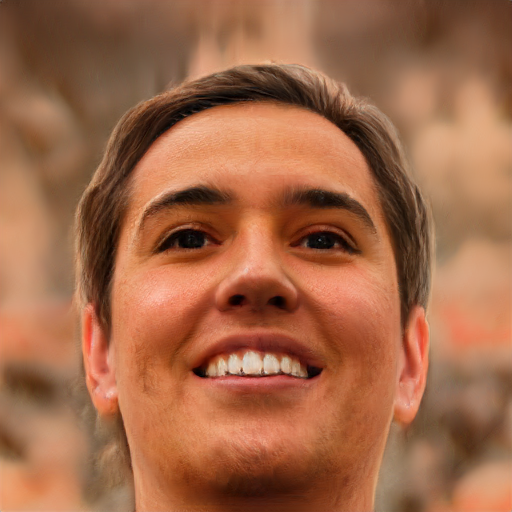}
        & \includegraphics[width=0.1666\linewidth]{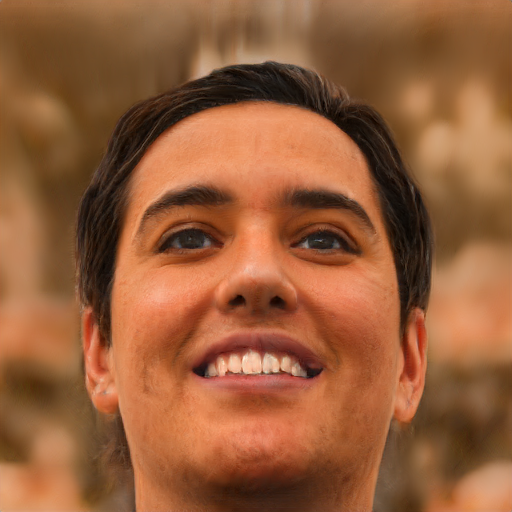}
       
        \\
        \includegraphics[width=0.1666\linewidth]{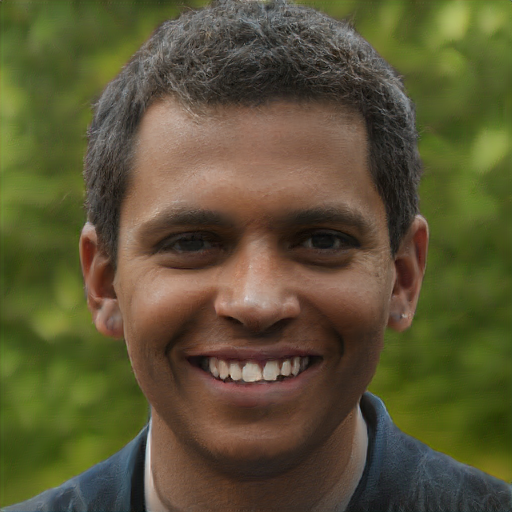}
        & \includegraphics[width=0.1666\linewidth]{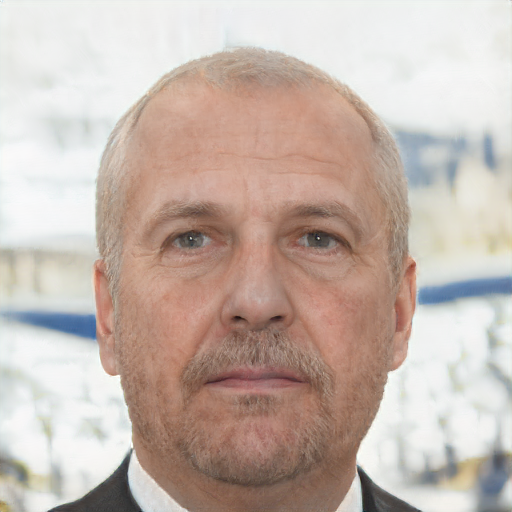}
        & \includegraphics[width=0.1666\linewidth]{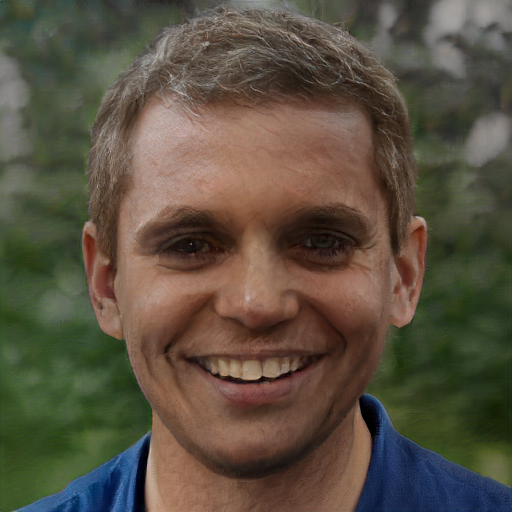}
        & \includegraphics[width=0.1666\linewidth]{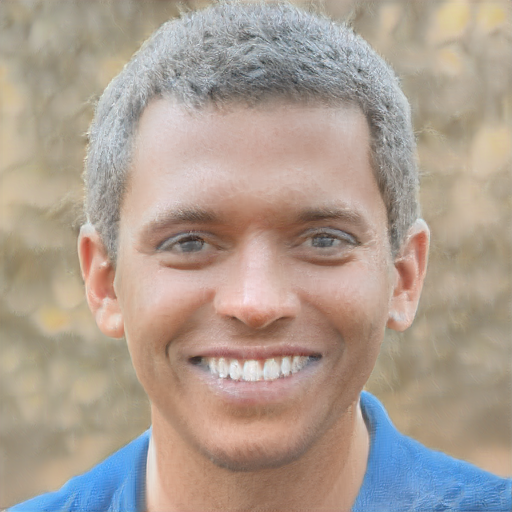}
        & \includegraphics[width=0.1666\linewidth]{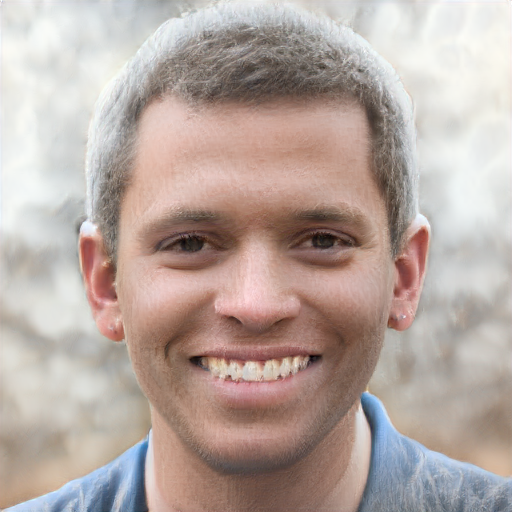}
        & \includegraphics[width=0.1666\linewidth]{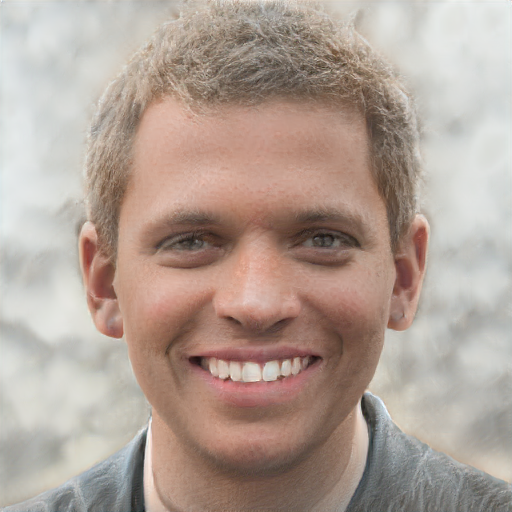}
        \\
        Geometry & Appearance & Baseline1 & Baseline2 & Baseline3 & Ours
    \end{tabular}
            \end{minipage}
    \end{tabular}
    
    \caption{The results of the ablation study. %
    (a) shows the ablation study for editing effects, while %
    (b) shows the ablation study for style transfer.}
    \label{fig:ablation_study}
\end{figure}
As for style transfer, our framework
can disentangle the geometry and appearance of a facial volume and generate {a} new facial volume by swapping geometry and/or appearance. {Thus, it can be used for 3D-aware style transfer, {as shown in Fig. \ref{fig:style-transfer-map} (please refer to the supplementary materials for more results)}.}
{For comparison, we perform style transfer between two real images.} From 
Fig. \ref{fig:comparison}(b), %
it is clear that during {style transfer}, SofGAN has obvious artifacts,
especially for {poses different from the original poses}. 
In contrast, our method not only successfully transfers the style but also preserves the geometry {accurately}.

\subsubsection*{Quantitative evaluations.}
We measure image quality with Frechet Inception Distance (FID) \cite{heusel2017fid} on the FFHQ dataset \cite{karras2019style}. Before fine-tuning, the FID of our implemented EG3D is $9.7$, while after fine-tuning the FID {increases to} %
$16.0$. %
Note {that} our fine-tuning needs less strict hardware requirements and shortened training time, {at the cost of slightly degraded quality}. %

\subsection{Ablation Study}
We conduct ablation studies to justify the necessity of each component in our framework.

For the ablation study of editing, 
we show the edited results of 
"Baseline1", which uses
a decoder 
to predict densities and semantic labels from
{the} original tri-planes. The color features are reused from the original fixed decoder $\Phi$. Thus, the geometry and appearance are not disentangled in this case.
We also test {"Baseline2", in which we use} %
an off-the-shelf 2D segmentation module \cite{yu2018bisenet} %
to parse generated images into semantic masks {for performing edits}.
Thus, the editing by optimization is performed on the fixed EG3D directly.
Besides, we also show the results of {"Baseline3", for which} %
we train a
decoder to predict semantic labels from
{the} original tri-planes, while reusing densities and color features %
from the fixed original decoder $\Phi$. Thus the semantic mask volume is not aligned with the facial volume.
From {Fig. \ref{fig:ablation_study}(a)}, %
it is clear that without the disentanglement ("Baseline1"), with {the} off-the-shelf segmentation module ("Baseline2"), or without the alignment between geometry and semantic masks ("Baseline3"), the editing effect{s fail} %
to be consistent with the modified semantic masks, such as the inaccurate hair length
and incomplete glasses in the first and second rows. %

{For the ablation study of style transfer, }we show the results by applying style-mixing {(denoted as "Baseline1")}{, since the latent codes of high-resolution blocks roughly control the appearance} (details can be found in the supplementary). 
Additionally,
we show the results of {the} generator decoupled by our method but without \(\mathcal{L}_{\text{Sim}}\) {(denoted as "Baseline2")}. We also give results with {the} histogram loss for a whole image instead of each label {(denoted as "Baseline3")}.
From the part (b) of Fig. \ref{fig:ablation_study}, it is clear that with the style-mixing ("Baseline1") or without $\mathcal{L}_{\text{Sim}}$ ("Baseline2"), the transferred style is inconsistent with the appearance reference.
Furthermore, when {the} histogram color loss {is} calculated for the whole image, 
Baseline3 generally captures the style but {is not} %
accurate at details. For example, %
the color of background diffuses to the region of hair.

\begin{figure}
    \renewcommand\tabcolsep{0.0pt}
    \renewcommand{\arraystretch}{0}
    \centering \small
    
    \begin{tabular}{c}
        \includegraphics[width=0.8\linewidth]{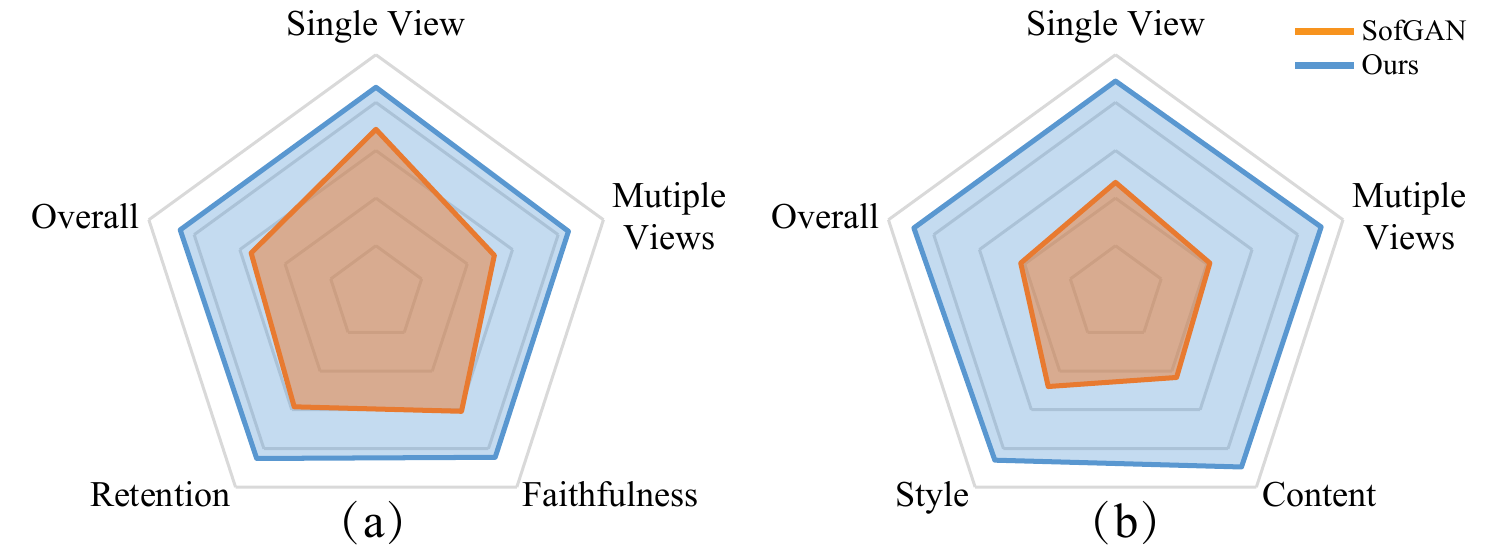}
    \end{tabular}
    
    \caption{{Radar plots of the average quality and faithfulness perception scores in terms of {the} similarity between the generated faces and the reference faces in appearance and geometry respectively, as well as the visual realism of the generated faces, in the single and multiple views. %
    (a) The comparison of editing with two methods: SofGAN \cite{chen2022sofgan} and ours. (b). the evaluation of style transfer with SofGAN and our method.}}
    \label{fig:perception-study}
\end{figure}

\subsection{Perception Study}
To evaluate the visual quality and the faithfulness of synthesized faces (i.e., the similarity to the geometry and appearance images), we conducted perception studies. 

{Specifically, we evaluate {the {performance} of} SofGAN and our {method} {for} %
editing and style transfer {using two respective} %
online questionnaires, %
in which each question is rated in a five-point Likert scale (1=strongly negative to 5=strongly positive). In the first {questionnaire}, %
we showed the original image and semantic mask, the modified semantic mask and results in the original view and {multiple other} %
views by {the} two methods, placed side by side in a random order to avoid bias. Each participant was asked to evaluate $15$ examples according to five criteria: the visual quality of generated face images in the original view, the visual quality of generated face images in {other} %
views, faithfulness to the changed regions, retention to the unchanged regions, and overall effects. In total, $25$ participants (7 female, 18 male, aged from 18 to 32, with normal vision) {without any special experience} participated in this study %
and we got $25$ (participants) $\times$ $75$ (questions) = $1875$ subjective evaluations for each method. We draw a radar plot (see Fig. \ref{fig:perception-study} {(a)}) in aspects of ``Single View'', ``Multiple Views'', ``Faithfulness'', ``Retention'', and ``Overall'' corresponding to the above-mentioned five criteria respectively. 
We found significant 
effects of our method for all five criteria based on scores in order: $4.32$, $4.23$, $4.23$, $4.26$ and $4.31$ over $3.44$, $2.60$, $3.04$, $2.92$, $2.75$ of SofGAN.
}

{In the second {questionnaire}, %
we presented users with 
the geometry and the appearance reference images, and {the} results by SofGAN and ours in the original view and multiple other views.
Each participant was asked to evaluate $15$ examples according to five criteria: the visual quality of synthesized face images in the original view, the visual quality of synthesized face images in other %
views, the maintenance of the geometry reference, the similarity to the appearance reference, and overall effects. In total, we got $25$ (participants same to the first questionnaire) $\times$ $75$ (questions) = $1875$ subjective evaluations for each method. %
Fig. \ref{fig:perception-study} {(b)} shows the statistics of these two methods, in which ``Single View'', ``Multiple Views'', ``Content'', ``Style'', and ``Overall'' correspond to the above-mentioned five criteria respectively. 
We get the values $4.45$, $4.52$, $4.48$, $4.30$ and $4.44$ compared to $2.33$, $2.07$, $2.16$, $2.40$, $2.09$ of SofGAN.}
{It is clear that our method achieved a significant 
improvement over {SofGAN}. %
}

{We have done the analysis of one-way ANOVA tests and paired t-tests for {the} %
two questionnaires with $p < 0.001$ for all {the tests}, %
which confirmed {our method significantly outperforms SofGAN}. %
For more detail, please refer to the supplementary materials.}

\begin{figure}
    \renewcommand\tabcolsep{0.0pt}
    \renewcommand{\arraystretch}{0}
    \centering \small
    
    \begin{tabular}{cccccc}
        Geo. &
        App. &
        Recon. & 
        Result &
        \multicolumn{2}{c}{Free-viewed Results}
        \\
        \includegraphics[width=0.1666\linewidth]{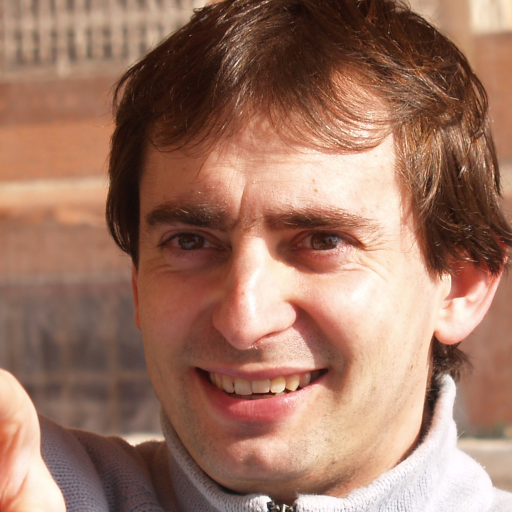} &
        \includegraphics[width=0.1666\linewidth]{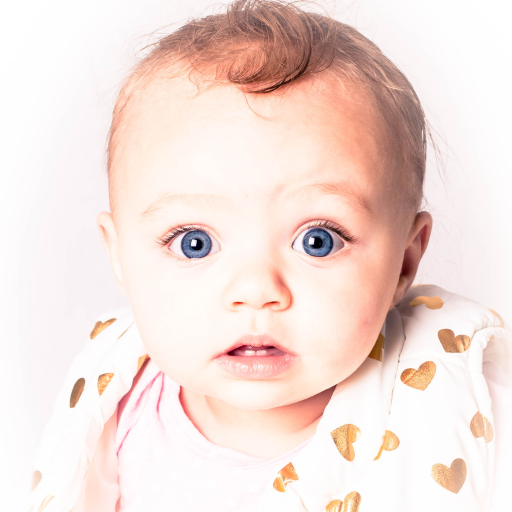} &
        \includegraphics[width=0.1666\linewidth]{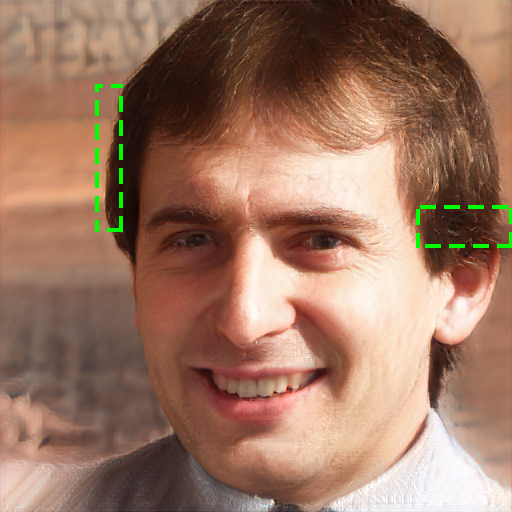} &
        \includegraphics[width=0.1666\linewidth]{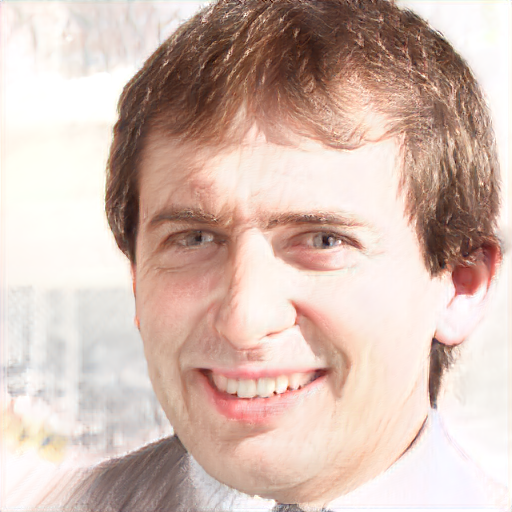} &
        \includegraphics[width=0.1666\linewidth]{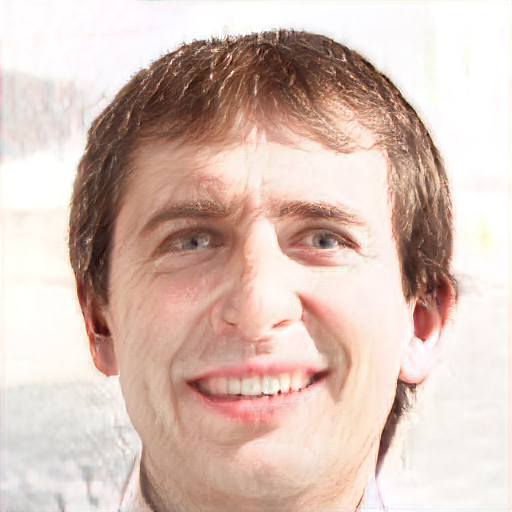} &
        \includegraphics[width=0.1666\linewidth]{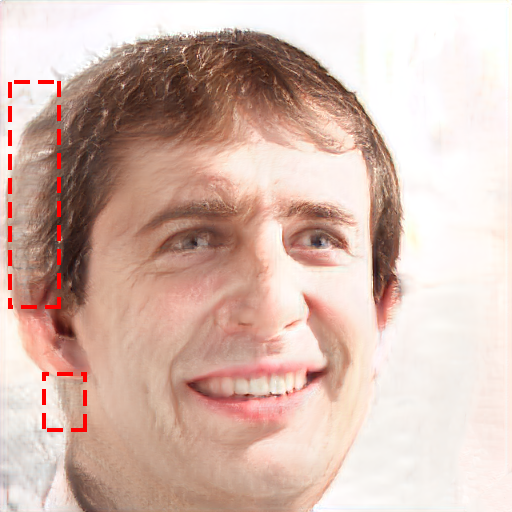}
    \end{tabular}
    \caption{%
    {A} failure case. {The red rectangles show the artifacts such as shadows or on-the-background neck textures. %
    The green rectangles show the lost details %
    such as the hair fringe through PTI. %
    Original images courtesy of pedronchi and Jane Gross.}
    }
    \label{fig:failing_case}
\end{figure}

\section{Conclusion, Limitations and Future Work}
{As the training for high-quality 3D-aware GANs is more and more time-consuming and resource-hungry, {approaches for} intuitive controlling {of} %
geometry and appearance based on pretrained 3D GANs should receive {more} %
attention. We believe that our method, a structured disentanglement framework based on the tri-plane representation, is a step forward in this direction, as we demonstrated our method can effectively edit the geometry through semantic masks and perform style transfer with fine-tuning while preserving the high speed and visual quality of {the} backbone.}
In the future, we are interested in implementing our framework based on the official EG3D and {evaluating} %
the impact of such a change.

{One limitation of this work is that due to the generation quality and inversion approach
of our model, our method has {artifacts {such as shadows} and} some ``billboards'' effects {at relatively large angles} and fail{s} to reach faithful and stereoscopic reconstruction (see Fig. \ref{fig:failing_case}). We speculate that they may come from our specific training process. 
Besides, in Fig. \ref{fig:failing_case}, rendered images are sometimes not realistic enough when performing {style transfer}. We attribute it to light and shadow{,} which our method cannot handle well.}
As future work, it would be useful to explore disentanglement of other attributes such as lighting to make our method more general.
Besides, semantic masks {have inevitable ambiguities while performing optimization {on the single view}, such as failing to control the gender {or holding unseen parts consistent}}.
Our method is capable of converting a real portrait image to its 3D avatar and {even possibly a talking head through editing guided by semantic masks}. %
Moreover, we can change its appearance while keeping its geometry unchanged. The editing may disturb the gender, and the style transfer may disturb the ethics of the original portrait. Therefore misusing our system potentially poses a societal threat, including ethics issues and fooling facial recognition system{s. We} %
do not condone using our work with the intent of spreading misinformation or tarnishing %
reputation. Thus, one should be careful to deploy this technology.
However, in case of misusing, existing methods (e.g., \cite{on-the-detection-of-digital-face-manipulation}) for detecting fake faces {might} alleviate this concern.

\begin{acks}

Thanks to Ms. Xiaohong Wang for her valuable comments and suggestions on the use of OneITLab.
This work was supported by grants from the National Natural Science Foundation of China (No. 62061136007, No. 62102403, and No. 61872440), Science and Technology Service Network Initiative of the Chinese Academy of Sciences (No. KFJ-STS-QYZD-2021-11-001), the Beijing Municipal Natural Science Foundation for Distinguished Young Scholars (No. JQ21013), ChinaPostdoctoral Science Foundation (No. 2022M713205), the Youth Innovation Promotion Association CAS. 
Hongbo Fu was supported by the Research Grants Council of HKSAR (No. 11212119), and the Centre for Applied Computing and Interactive Media (ACIM) of School of Creative Media, City University of Hong Kong.
\end{acks}

\bibliographystyle{ACM-Reference-Format}
\bibliography{bibliography}

\end{document}